\documentclass[twocolumn,tighten,twocolappendix]{aastex63}
\usepackage{amsmath}
\usepackage{enumerate}

\usepackage{txfonts}
\usepackage{booktabs}
\usepackage{multirow}
\usepackage{wrapfig}
\newcommand{\kms}[1]{{km\,s$^{-1}$#1}}

\shorttitle{{The strongest  magnetic field in a light bridge}}
\shortauthors{Castellanos Dur\'{a}n, Lagg, Solanki \& van Noort}
\received{}
\revised{}
\accepted{}
\submitjournal{ApJ}

\begin{document} 

\title{{Detection of the strongest magnetic field in a sunspot light bridge} }
\correspondingauthor{J. Sebasti\'an Castellanos Dur\'an}
\email{castellanos@mps.mpg.de}

\author[0000-0003-4319-2009]{J. S. Castellanos Dur\'an}
\affiliation{Max Planck Institute for Solar System Research, Justus-von-Liebig-Weg 3, 37077 G\"ottingen, Germany}

\author[0000-0003-1459-7074]{Andreas Lagg}
\affiliation{Max Planck Institute for Solar System Research, Justus-von-Liebig-Weg 3, 37077 G\"ottingen, Germany}

\author[0000-0002-3418-8449]{Sami K. Solanki}
\affiliation{Max Planck Institute for Solar System Research, Justus-von-Liebig-Weg 3, 37077 G\"ottingen, Germany}
\affiliation{School of Space Research, Kyung Hee University, Yongin, 446-101 Gyeonggi, Republic of Korea.}

\author{Michiel van Noort}
\affiliation{Max Planck Institute for Solar System Research, Justus-von-Liebig-Weg 3, 37077 G\"ottingen, Germany}

  \begin{abstract}
Traditionally, the strongest magnetic fields on the Sun have been measured in sunspot umbrae. More recently, however,  much stronger fields have been measured at the ends of penumbral filaments carrying the Evershed and counter-Evershed flows. Superstrong fields have also been reported within a light bridge separating two umbrae of opposite polarities. We aim to accurately determine the strengths of the strongest fields in a light bridge using an advanced inversion technique and to investigate their detailed structure. We analyze observations from the spectropolarimeter on board the Hinode spacecraft of the active region AR 11967. The thermodynamic and magnetic configurations are obtained by inverting the Stokes profiles using an inversion scheme that allows multiple height nodes. Both the traditional 1D inversion technique and the so-called 2D coupled inversions, which take into account the point spread function of the Hinode telescope, are used. We find a compact structure with an area of 32.7 arcsec$^2$ within a bipolar light bridge with field strengths exceeding 5\,kG, confirming the strong fields in this light bridge reported in the literature. Two regions associated with downflows of $\sim$5\,km\,s$^{-1}$ harbor field strengths larger than 6.5\,kG, covering a total area of 2.97 arcsec$^2$. The maximum field strength found is 8.2\,kG, which is the largest ever observed field in a bipolar light bridge up to now. 
\end{abstract}


\section{Introduction}\label{sec:intro}

Sunspots are the most striking magnetic features at the solar surface. Their darkness relative to their surroundings is caused by the strong magnetic fields they harbor, which suppress convective energy transport \citep[e.g.,][]{Solanki2003,Rempel2011LRSP}. It is unclear, however, how strong the field in sunspots can be and, in particular, where in sunspots the strongest fields are found.

\citet{Livingston2006SoPh} investigated archives from three different observatories that gathered data from 1917 to 2004, finding 55 spots with magnetic field strengths larger than 4\,kG, with the strongest field being 6.1\,kG. These authors found these strong fields inside the darkest parts of the sunspot umbrae. Equally strong or even stronger fields have been reported in highly sheared regions, where two active regions (ARs) with different polarities collide \citep[e.g.,][]{Zirin1993b,Wang2018RNAAS}. Thus, \citet{Wang2018RNAAS} reported fields of 5.57\,kG measured directly from the Zeeman splitting of the \ion{Fe}{1} 1.5648 $\mu$m spectral line. \citet{Okamoto2018ApJ} obtained magnetic field strengths up to 6.25\,kG inside the light bridge of AR 11967 by fitting the observed Stokes profiles assuming a Milne-Eddington (ME) type atmosphere. These authors stated that their measurements correspond to the strongest field ever reported directly deduced from  Zeeman splitting in Stokes $I$. 

Recently two locations harboring very strong fields were detected in two different parts of the penumbra \citep{vanNoort2013A&A,Siu-Tapia2017A&A,Siutapia2019}. The first detected region is located at endpoints of penumbral filaments. The strong fields are related to strong and often supersonic downflows \citep{vanNoort2013A&A}. These authors obtained field strengths reaching up to 7.25\,kG in the deepest layers accessible to observations at such locations with velocities up to 20\,\kms{}. The second reported region is a peculiar piece of penumbra, displaying inward motion at the boundary between the penumbra and umbra \citep{Siu-Tapia2017A&A}. These inward flows are also known as counter-Evershed flow owing to the reversal of the direction compared to the classical Evershed flow \citep{Evershed1909,Siu-Tapia2018ApJ}. \citet{Siu-Tapia2017A&A} found an area of more than 5.1 arcsec$^2$ with a field strength larger than 7\,kG, with maximum values of $\sim$8.3\,kG at the strong downflow regions bordering the umbra. However, these authors stated that their fits to the complex Stokes profiles were not as good as in other places of the analyzed AR and therefore excluded them from their analysis in 2017. In recent work, \citet{Siutapia2019} studied the likelihood of these complex Stokes profiles to be produced either by strong fields or by a multicomponent atmosphere within the resolution element. These authors, based on Bayesian analysis, concluded that the strong-field scenario is the most likely explanation for their observations. It is worth noting that both strong fields in the inner and outer parts of the sunspot were both associated with superfast downflows. In this paper, we present a third location in which to find strong magnetic fields: a bipolar light bridge. However, the strong fields presented here differ from previous ones in their association with slow downflow velocities.

\begin{figure*}[thp]
 \begin{center}
\includegraphics[width=0.49\textwidth]{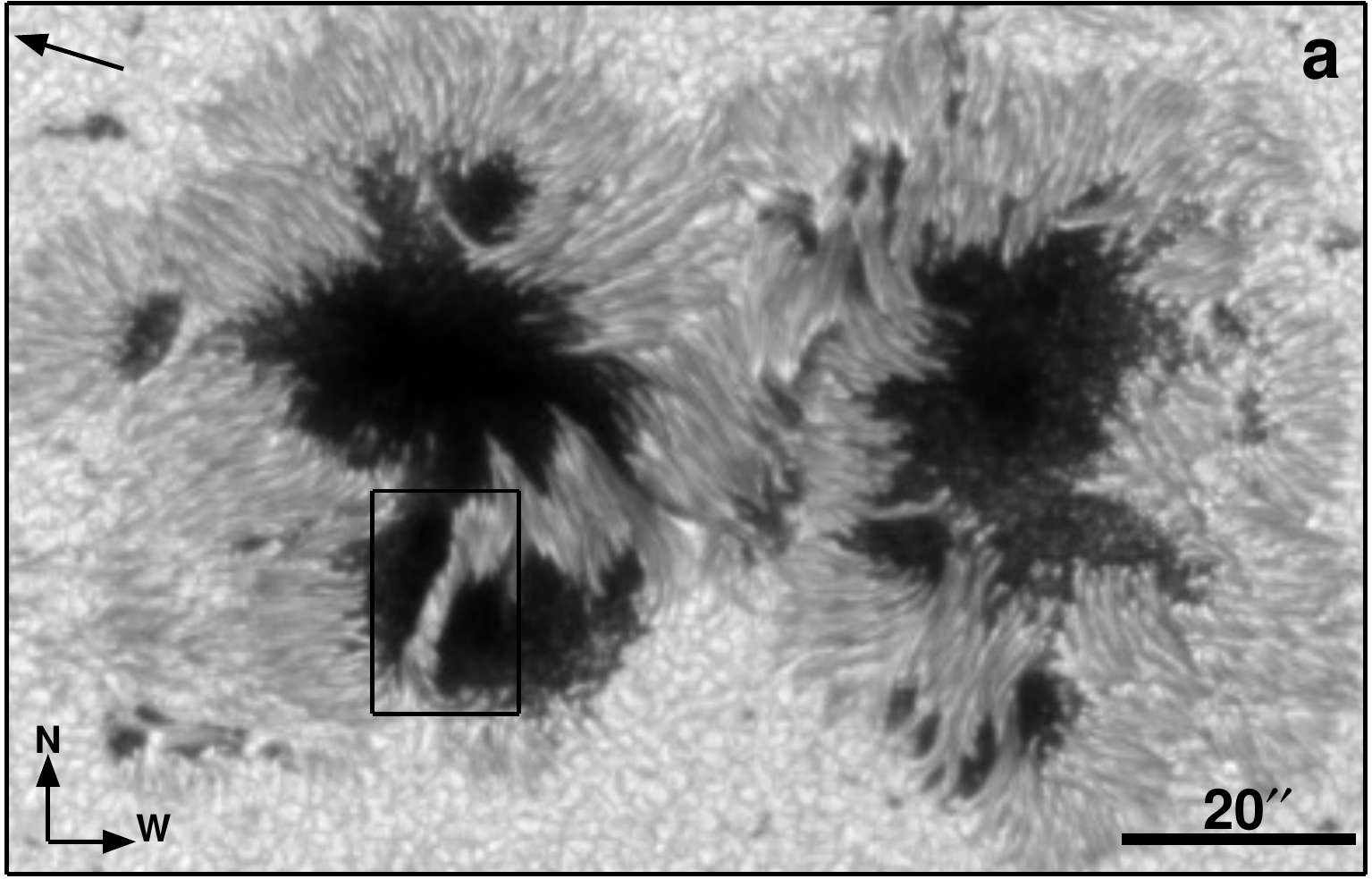}
\includegraphics[width=0.49\textwidth]{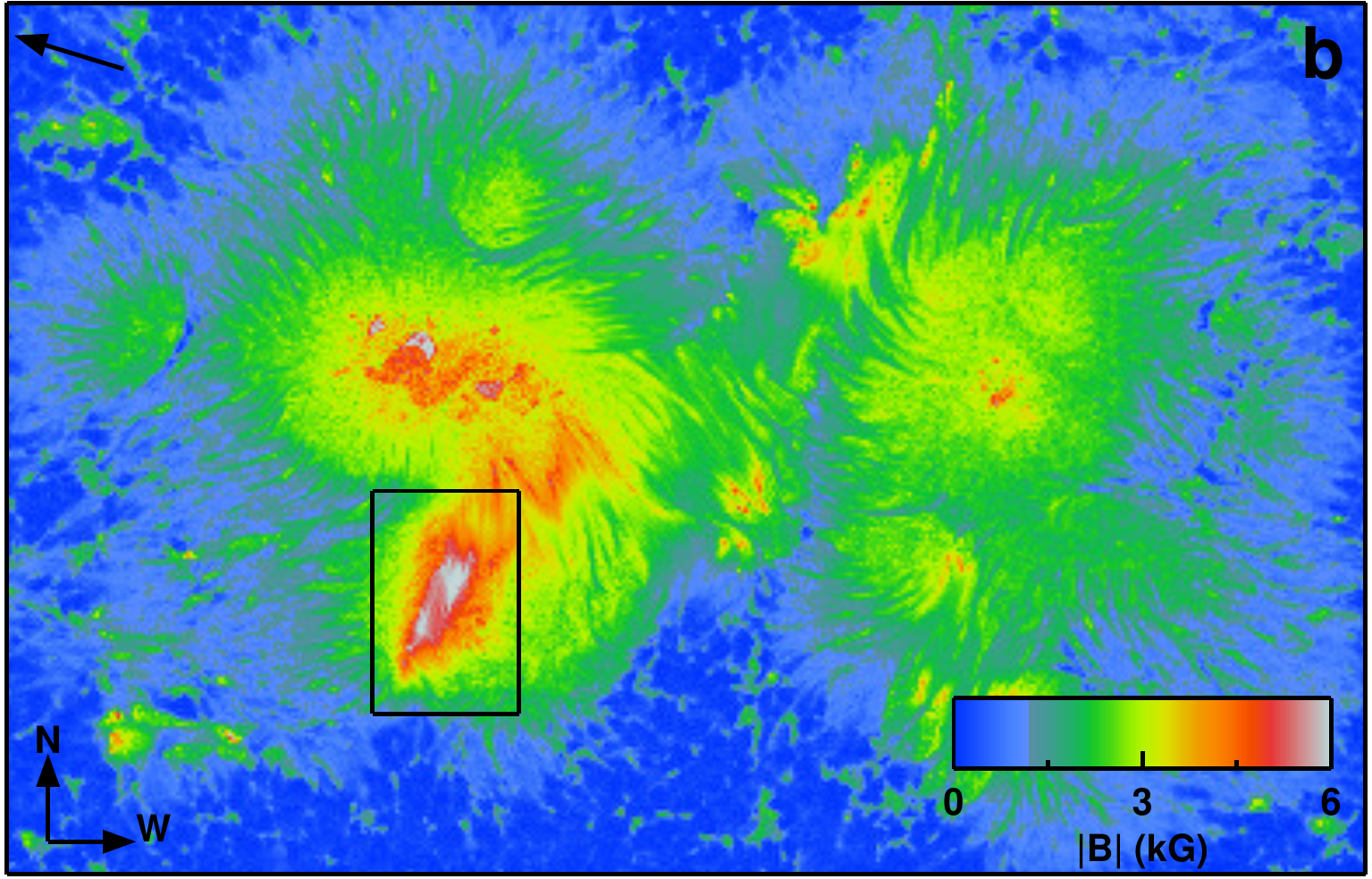}
\includegraphics[width=0.49\textwidth]{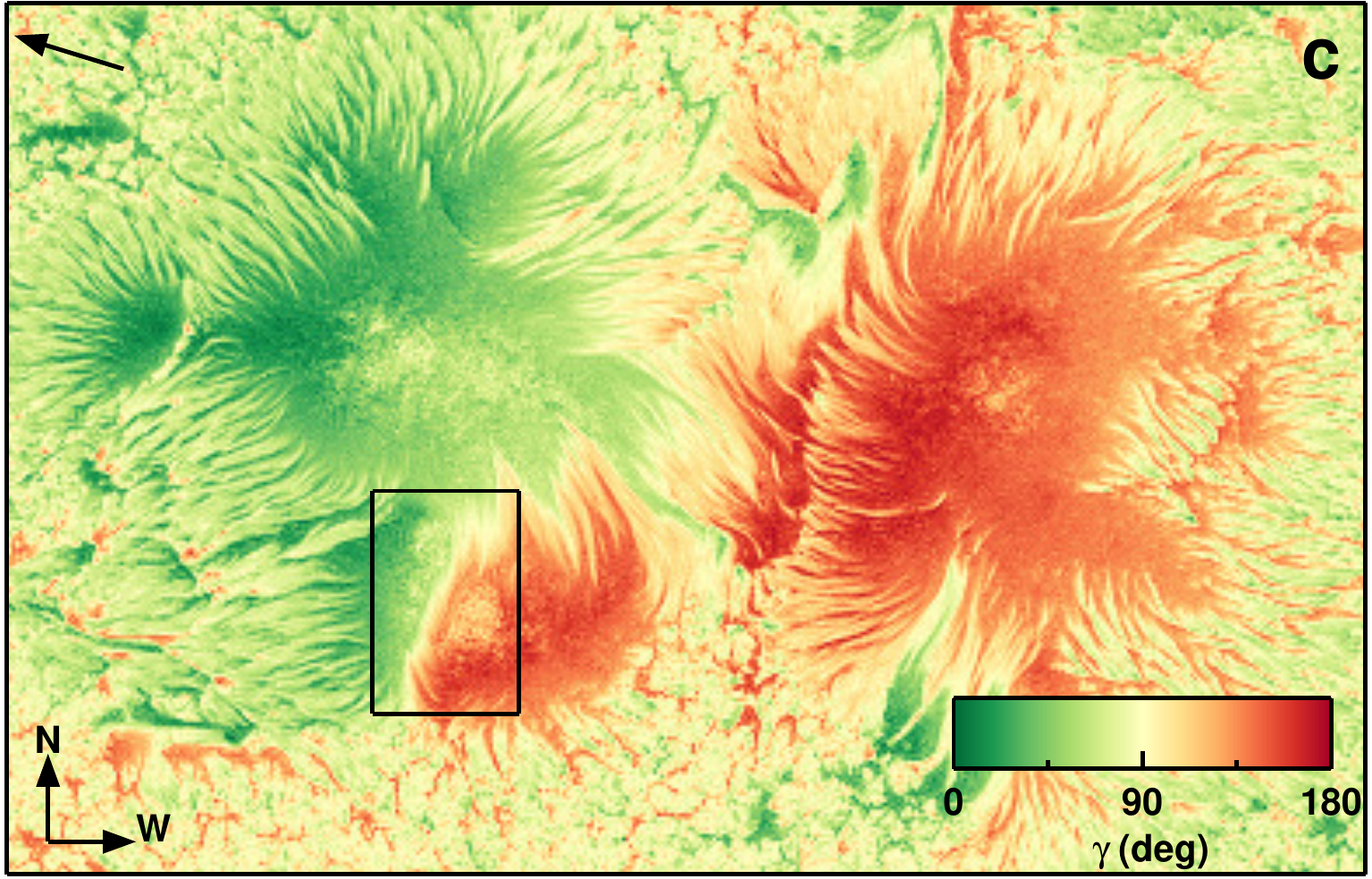}
\includegraphics[width=0.49\textwidth]{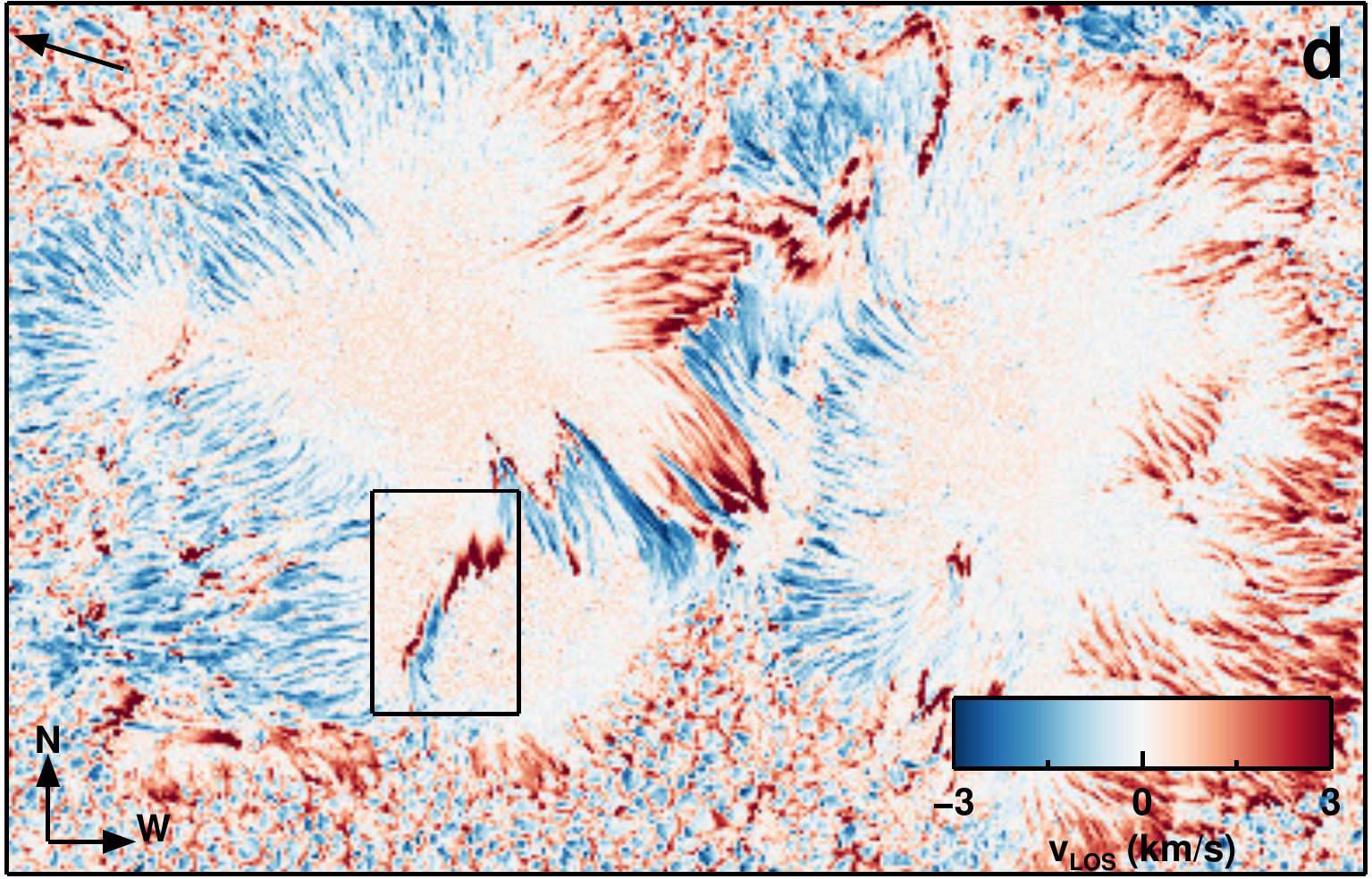}

 \caption{
 {(a) Continuum image of AR 11967 taken by Hinode/SOT-SP on 2019 February 4 at 19\,UT. The black rectangles include the light bridge containing the superstrong magnetic field.
 (b) Maps of the magnetic field strength, (c) inclination, and (d) velocity with respect to the line of sight, all at $\log\tau\!=\!-0.8$ and all obtained by the 2D inversion. See main text for details.  The azimuth of the magnetic field at the region of interest is presented in Figure~\ref{fig:shear}a. {The location of the AR is $16^{\circ}$\,W, $8^{\circ}$\,S.}}
 }\label{fig:ar}
 \end{center}
 \end{figure*}

Light bridges are elongated intrusions that appear in the umbra of sunspots \citep[e.g.,][]{Leka1997}. \citet{Lagg2014A&A} stated that light bridges can usually be divided into three categories depending on their size and brightness:   faint light bridges \citep{Lites1991}, {\it strong} light bridges \citep{Sobotka1993}, and granular light bridges \citep{Vazquez1973,RouppevanderVoort2010}. These categories share the property of a rather weak field strength on the order of hundreds of gauss or even lower in their deepest observable layers \citep[e.g.,][]{Lagg2014A&A, Toriumi2015}. However, there is another type of light bridge, not mentioned in \citet{Lagg2014A&A}, one separating umbrae of opposite polarities, i.e. light bridges in delta-spots. The only possibility for such a bipolar configuration of sunspot umbrae to develop is the convergence of two ARs. Such light bridges appear along the polarity inversion line (PIL) and in some cases have been found to harbor strong fields of the order of $\gtrsim$4\,kG \citep[e.g.,][]{Tanaka1991SoPh, Zirin1993b, Livingston2006SoPh,Jaeggli2016ApJ,Okamoto2018ApJ,Wang2018RNAAS}.

In this paper, we have analyzed the same observations as \citet{Okamoto2018ApJ}, but with more sophisticated tools to extract more of the information encoded in the sunlight. We interpret the observations by solving the radiative transfer equations assuming a height-stratified atmosphere and taking into account the point spread function (PSF) of the telescope. This type of inversion builds on more realistic assumptions than the simplistic, ME-type atmospheres, allowing us to determine the magnetic properties of these light bridges and in particular of the strong fields they harbor in greater detail.

\section{Observations} \label{sec:obs}

The main data set we use to infer the magnetic field information at the light bridge was recorded by Hinode/Spectro-Polarimeter (SP). We also employ Hinode/BFI filtergrams and Helioseismic and Magnetic Imager (HMI) data to follow the evolution of the AR as it passed over the solar disk. 

We studied 32 scans taken by Hinode/SP of AR 11967. These scans were taken from 2014 February 1 to February 6. In this article we focus our analysis on the scan starting at 19:00~UT on 2014 February 4, when the AR was located at $16^{\circ}$\,W, $8^{\circ}$\,S. The selected scan is the same as analyzed by \citet{Okamoto2018ApJ} and is the one in which the strongest magnetic field is found. The spectropolarimetric data were obtained with the Solar Optical Telescope \citep[SOT;][]{Tsuneta2008}, specifically with the SP \citep{Ichimoto2008SoPh} on board the Hinode satellite \citep{Kosugi2007}. Hinode/SP measures the full Stokes vector of the \ion{Fe}{1} spectral lines at 6301.5\,\AA{} and 6302.5\,\AA{}, with a spectral sampling of 21.5\,m\AA{}. The plate scale along the slit is 0\farcs32 pixel$^{-1}$, and 0\farcs29  pixel$^{-1}$ along the scan direction (=fast scan mode).  {Figure~\ref{fig:ar} shows the continuum image and the maps of magnetic field strength, inclination, and line-of-sight velocity of AR 11967. The black boxes on the left side of each panel of Figure~\ref{fig:ar} mark the region of interest that harbored strong magnetic fields.}

The data are calibrated using the standard reduction tools \citep{Lites2013SoPh}. These routines account for spurious continuum polarization (SCP). However, for those profiles with extreme Zeeman splitting, the wing of the {\ion{Fe}{1}} line is extremely broad and therefore affects the result of the standard calibration procedures. As a consequence, an offset in Stokes $Q$ and $U$ is detected. This effect can be easily seen, for example, in Figure 1 of \citet{Okamoto2018ApJ}, where the regions with strong fields show clearly nonzero Stokes $Q$ and $U$ intensities in the continuum, where the signal should be zero. This spurious signal is on average $\sim$0.5\% but can be as high as 2\% in the linear polarization. Since this effect only occurs in a few pixels, we correct it post facto by simply subtracting this offset. In the case of \citet{Okamoto2018ApJ}, the SCP does not affect their conclusions significantly, as these authors assumed an ME-type atmosphere. However, for a height-dependent inversion, like the one used for this study, the SCP strongly affects the information retrieved at lower heights (see next section), since this information is predominantly contained in the wings of the spectral lines.

We followed the temporal evolution of the AR 11967 using data from the HMI \citep{Scherrer2012} on board the Solar Dynamics Observatory \citep[SDO;][]{Pesnell2012}. We took continuum intensity images and magnetograms to track the AR from a longitude of -45$^{\circ}$ to +45$^{\circ}$ every 30 minutes. This interval ranges from January 31 at 02:00~UT to February 6 at 22:00~UT. In addition, we used a full 45\,s cadence lasting for 10 hr around the Hinode's scan starting on February 4 at 17:00~UT. The continuum images are enhanced using the neuronal network \texttt{Enhance} \citep{DiazBaso2017}. {In addition, we use Hinode/BFI observations starting on February 4 at 00:00\,UT and lasting for 26 hr. We focus on the filtergrams of the {\ion{Ca}{2}}\,h at $3968.5$\,\AA{} with 1-minute cadence and the $G$-band filtergrams around $3883.5$\,\AA{} with 10-minute cadence. Note that there are some gaps in the data that depend on the observing mode. Three videos are provided as online material. Snapshots of each video are described in Figures \ref{fig:video1}-\ref{fig:video3} in Appendix \ref{sec:videos}.}

\begin{figure*}[tbhp]
 \begin{center}
 \includegraphics[width=.95\textwidth]{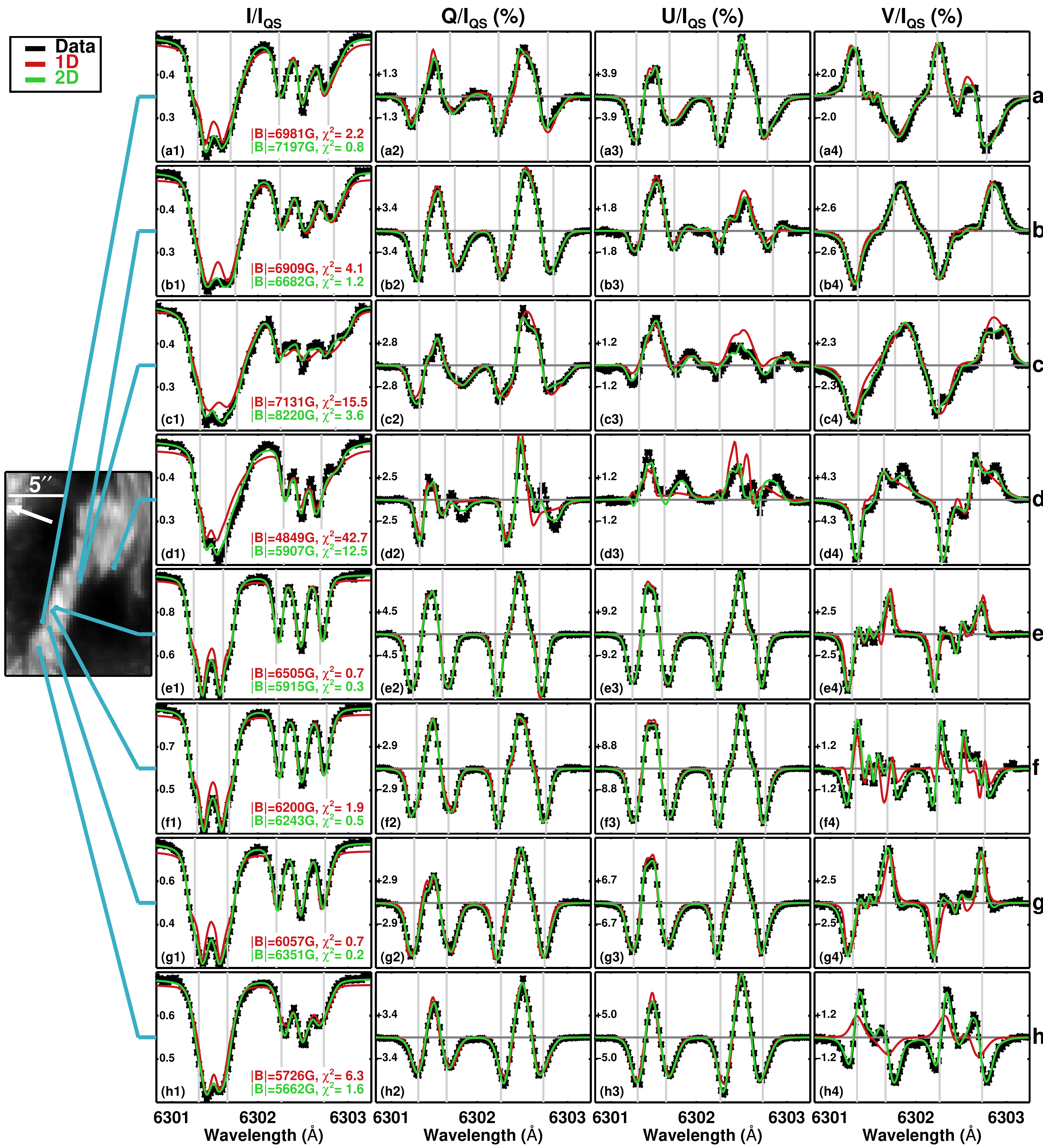}
 \caption{Comparison of the fits of the Stokes profiles from a 1D and spatially coupled 2D inversions at seven representative pixels within the light bridge. The leftmost panel shows a blowup of the light bridge (scene within the black rectangle in Figure~\ref{fig:ar}), with the blue lines indicating to which pixel each profile displayed in the other panels refers. The remaining subplots show from left to right the four Stokes profiles, with the observations being represented by black crosses. The field strengths given in the Stokes $I$ panels refer to $\log\tau\!=\!0$. The best fits resulting from the inversions are presented in red (1D inversion) and green (2D inversion). Vertical gray lines show the Zeeman splitting ($\Delta\lambda_{|\vec{B}|}\!=\!4.67\times10^{-13}g_{\rm eff}\lambda_0|\vec{B}|$) by a magnetic field with the same amplitude as the retrieved value of the 2D inversion at $\log\tau\!=\!-0.8$. These lines are Doppler-shifted by $v_{\text{LOS}}$ at the same height. Each row exemplifies the following scenarios: {(a) $|\vec{B}|_{\rm 1D}\!\approx\!|\vec{B}|_{\rm 2D}\approx7$\,kG}, (b) $|\vec{B}|_{\rm 1D}\!>\!|\vec{B}|_{\rm 2D}$, (c) the pixel with the largest field, (d) possiblle multicomponent atmosphere within the resolution element (see Figures \ref{fig:2comp2d}-\ref{fig:2comp1d} for further analysis),
 (e) region next to the PIL, (f) region at the PIL ($\gamma\approx90^{\circ}$), (g) region with $|\vec{B}|\!>\!6$\,kG at $\log\tau\!=\!-0.8$ for both inversions, and (h) pixel where the 1D inversion failed. To better explore the quality of the fit rather than the intensity of the profiles, each panel has been scaled to the maximum and minimum of the observed profile normalized to the HSRA continuum, while the Stokes $Q$, $U$, $V$ are displayed symmetrically with respect to the zero polarization (horizontal gray line). }
 \label{fig:inv}
 \end{center}
 \end{figure*}

\section{Inversions} \label{sec:inversions}

 \begin{figure}[tbh]
 \begin{center}
 \includegraphics[width=.48\textwidth]{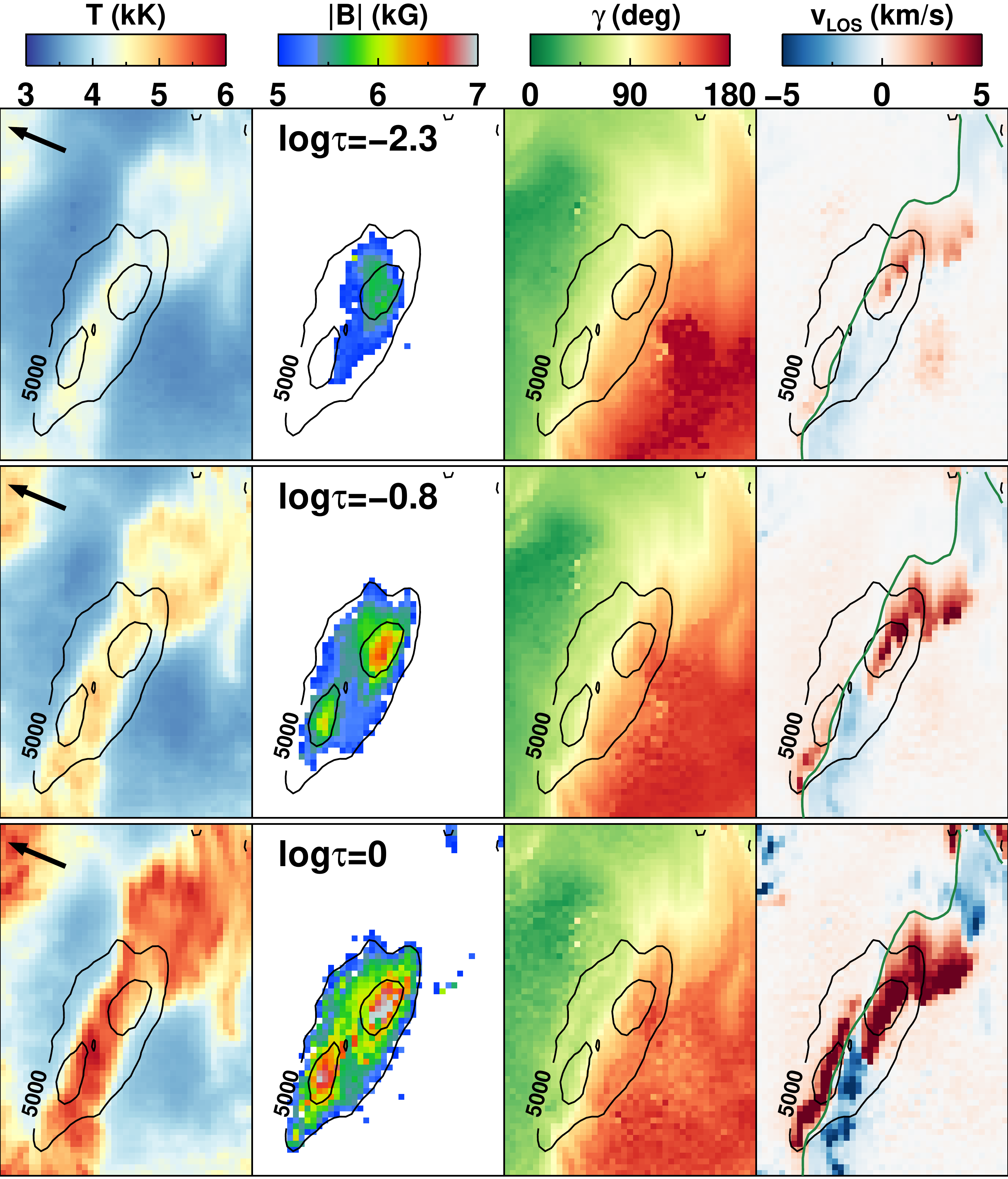}
 \caption{Maps of atmospheric parameters obtained with the 1D inversions. Columns show, from left to right, temperature, {magnetic field strength}, and the inclination and velocity relative to the line of sight. Each row denotes different optical depths: $\log\tau\!=\!-2.3$ (top), $\log\tau\!=\!-0.8$ (middle), and $\log\tau\!=\!0$ (bottom). The arrows in the leftmost column point to the solar disk center. The color bar of a column applies to all rows of that column. Overplotted on the images are field strength contour levels at 5.0 and 6.1\,kG at $\log\tau\!=\!0$. The green line in the last column traces the PIL at each optical depth. The azimuth at the light bridge is presented in Figure~\ref{fig:shear}.
 }\label{fig:atm1d}
 \end{center}
 \end{figure}
 \begin{figure}[tbh]
 \begin{center}
 \includegraphics[width=.48\textwidth]{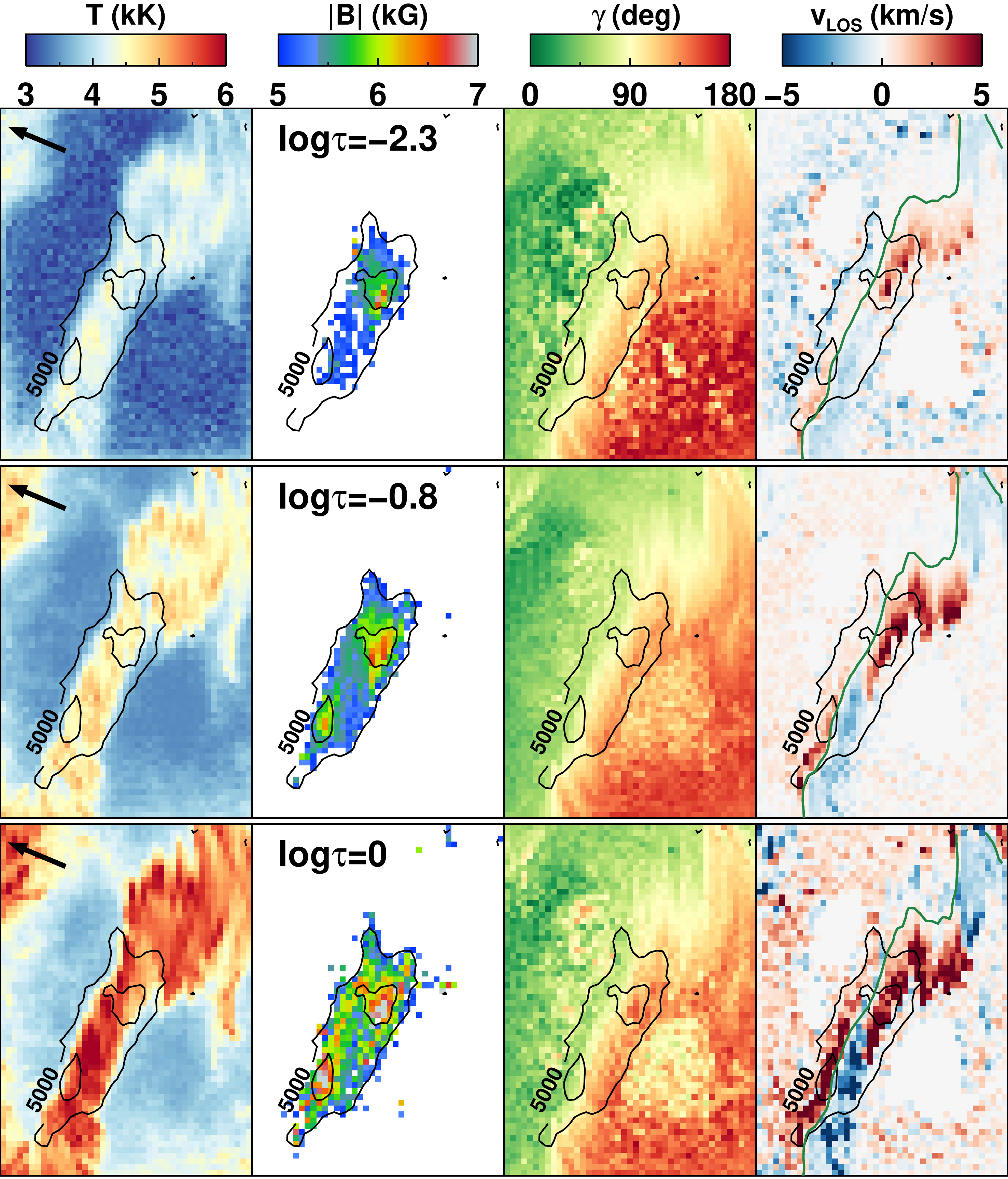}
 \caption{Atmospheric parameters retrieved from the 2D inversions. The layout is the same as in Figure~\ref{fig:atm1d}.}\label{fig:atm2d}
 \end{center}
 \end{figure}

In an optical system, the PSF determines how an observed point source is imaged on the detector plane. Inversely, the PSF can also be used to calculate how much of the information within a resolution element (pixel) actually comes from the surroundings. For a diffraction-limited instrument, more than 80\% of the photons originating from a point source on the solar surface are distributed to the neighboring pixels, assuming a spatial sampling at the Nyquist frequency. {In the current dataset, however, the data are undersampled by a factor of almost two, limiting the expected contamination of each pixel by the surroundings to less than 25\%.}

Solar magnetometry is based on measuring the degree of polarization of sunlight to infer the magnetic field properties in the Sun. Light coming from two different magnetic structures on the solar surface will have, in principle, two different degrees of polarization. After passing the optical system, the measurements of light from these two objects are blurred owing to the PSF, thus mixing the observed degree of polarization from both objects. On top of that, if we take into account the central obscuration of Hinode/SP that introduces a more complex PSF into the system, then properly addressing the PSF while analyzing spectropolarimetric data is clearly necessary.

We carried out two types of stratified inversions using the SPINOR code \citep[][which relies on the STOPRO routines \citep{Solanki1987PhDT} to do the polarized radiative transfer]{Frutiger2000}: the traditional pixel-by-pixel (1D) approach, and a spatially coupled scheme that takes into account the PSF of the instrument \citep[hereafter denoted as 2D inversion following][]{vanNoort2012A&A}. Three nodes were used, located at $\log\tau\!=-2.3$,\,-0.8, and 0.0, where $\tau$ is the continuum optical depth at $5000$\,\AA{}, for all relevant atmospheric parameters, which serve as free parameters used to obtain a fit to the data. This choice spans the formation heights of the {\ion{Fe}{1}} lines at 6300\,\AA{}, which lie, depending on the atmospheric parameter, in the range of $\log\tau\!=\![+0.1, -2.7]$ assuming a standard VAL-C model \citep{Vernazza81}.  {The locations of the nodes were optimized to obtain the global minimum for the entire map of the fits. This is important for the 2D inversions, since the information from different pixels is coupled. As a consequence of this global approach, there might be locations where the node placement is locally not optimal. In addition, we performed tests by placing the nodes at different optical depths and repeating the inversions (see Table~\ref{tab:nodes} in the appendix~\ref{sec:nodesposition}). For example, we shifted the bottom node even below the $\log\tau\!=\!0$. We also tried setting two nodes for the $v_{\text{LOS}}$. However, all our experiments clearly maintained the strong-field character of the light bridge (see appendix ~\ref{sec:nodesposition} for details). Therefore, the choice of node location does not strongly affect the reported results.}

The free parameters of the inversion are the temperature ($T$), magnetic field strength ($|\Vec{B}|$), inclination ($\gamma$), azimuth ($\phi$), line-of-sight velocity ($v_{\rm LOS}$) and microturbulence ($v_{\rm micro}$). It is worth noting that both inversion schemes have the same number of free parameters. 
A Gaussian kernel with an FWHM of 24.3\,m\AA{} is used to account for the spectral instrumental broadening. No further parameter is needed to justify the broadening of the line, such as a macroturbulence.  We also verified that for the {\ion{Fe}{1}} line pair at 6300\,\AA{} and a field strength of 6000 G, i.e. close to the maximum value reported by \citet{Wang2018RNAAS}, we are still far away from the Paschen-Back regime, which was ignored in our analysis.

Examples of the fits to the observed Stokes profiles within the light bridge are shown in Figure~\ref{fig:inv}.  Black crosses indicate the observed data, while the best-fit profiles resulting from the 1D classical inversion scheme and the 2D inversion are represented by the red and green lines, respectively. {Vertical gray lines show the Zeeman splitting by a magnetic field $|\vec{B}|$, given by $\Delta\lambda_{|\vec{B}|}\!=\!4.67\times10^{-13}g_{\rm eff}\lambda_0|\vec{B}|$, where $g_{\rm eff}$ is the effective Land\'e factor of the transition with a wavelength $\lambda_0$. We used the magnetic field retrieved by the 2D at $\log\tau\!=\!-0.8$}. For all the plotted profiles, the fit by the 2D inversion scheme is better both visually and quantitatively (in the sense that $\chi^2_{\rm 1D}\!\gtrsim\!\chi^2_{\rm 2D}$). 
Figure~\ref{fig:inv} corroborates how well the 2D inversions fit the observed Stokes profiles, even in extremely complex cases, such as those in panels (c), (e), (f), and (h). Only in panel (d) is the fit less perfect, as evinced by the fact that the  reduced  $\chi^{2}$ for both inversion schemes is significantly larger than unity. As we shall discuss later, the possibility of the existence of a second atmospheric component within the resolution element can be excluded (Section~\ref{sec:2comp}). For all other profiles, the complexity can be fully explained by the light contributed to that pixel by neighboring pixels, which often have quite different profile shapes, so that the observed profile looks quite complex, although the atmospheric structure in that particular pixel may be relatively simple. Thus, these very complex profiles can be reproduced by simple one-component models if the stray light from the other pixels is properly taken into account. 

Magneto-optical effects can be excluded as a source of the complex profiles observed near the PIL.  These effects contribute negatively to the polarization degree in the absorption matrix, resulting in a reversal at wavelengths where the core of the line is. However, this sign reversal is not observed (see, e.g., rows (e) and (g) in Figure~\ref{fig:inv}).  

\section{Results} \label{sec:results}
\subsection{Atmospheric conditions at the light bridge}

Following the inversion schemes described in section \ref{sec:inversions}, the atmospheric maps obtained by 1D and 2D inversions are presented in Figures \ref{fig:atm1d} and \ref{fig:atm2d}, respectively. Columns give temperature, magnetic field strength, inclination, and line-of-sight velocity at the three optical depth nodes. The optical depth for each row is indicated in the field strength column. The levels of the field strength contours are 5 and 6.1\,kG, \citep[the latter corresponds to the largest field reported by][]{Livingston2006SoPh}. 
Clear evidence of the smearing by the PSF can be seen in the size of the {area covered by the magnetic field}. In the 1D inversion the atmospheric maps appear more blurred, with  the contours for the field $>$5\,kG being roundish and extending far outside the light bridge. In contrast, the 2D inversion confines the strong fields on the light bridge, and the 5\,kG contour nicely follows the light bridge shape. 

The PIL passes through the light bridge and separates not only the magnetic polarities but also, in the southern part of the light bridge, the negative and positive Doppler velocities (green lines on the fourth column in Figures \ref{fig:atm1d} and \ref{fig:atm2d}). An area of 32.7 arcsec$^2$ harbors fields larger than 5\,kG. The strong fields occur in the bright region such as a light bridge, and not in the dark umbra. Table~\ref{tab:area} lists the number of pixels for both inversions with fields larger than the thresholds from 5 to 7.5\,kG at $\log\tau\!=\!0$ and $\log\tau\!=\!-0.8$ in parentheses.

\begin{table}[tbhp]
\centering{}%
\begin{tabular}{ r  l l  ll}
\toprule[1.5pt]
\multicolumn{1}{c}{}&\multicolumn{2}{c}{1D inversions}&\multicolumn{2}{c}{2D inversions}\tabularnewline
\cmidrule(lr{3pt}){2-3}\cmidrule(lr{3pt}){4-5}
$|B|>$& Pixels & arcsec$^2$ & Pixels & arcsec$^2$  \tabularnewline
\midrule[1.5pt]
5.0\,kG& 417 (333) & 38.7 ({30.9}) & 352 (287) & 32.7 (26.6) \tabularnewline
6.0\,kG& 123 (29)  & 11.4 (2.7) & 105 (24) & 9.7 (2.2)\tabularnewline
6.5\,kG& 38 (3)    & 3.5 (0.3) &32 (5) & 3.0 (0.5)\tabularnewline
7.0\,kG& 6 (0)     & 0.6 (0)& 9 (0)& 0.8 (0)\tabularnewline
7.5\,kG& 0 (0)     & 0 (0) & 3 (0)& 0.3 (0)\tabularnewline 
\bottomrule[1.5pt]
\end{tabular}\caption{Area covered by the strong fields at different thresholds at $\log\tau\!=\!0$ and, in parenthesis, at $\log\tau\!=\!-0.8$. The pixel size is 0.29\arcsec$\times$0.32\arcsec.}\label{tab:area}
\end{table} 

Both 1D and 2D inversions show two regions with fields stronger than 6.5\,kG mainly at $\log\tau\!=\!0.0$ and $\log\tau\!=\!-0.8$, but in a few pixels also at $\log\tau\!=\!-2.3$. These two regions are associated with downflows observed at all three nodes. The velocity is higher in deeper layers, and the mean velocity is around 5\,\kms{} at $\log\tau\!=\!0$, which is in the subsonic regime. Inside these regions, the strongest magnetic fields at $\log\tau\!=\!0$ are $\sim$7.3\,kG (1D) and $\sim$8.2\,kG (2D), while at $\log\tau\!=\!-0.8$ the largest field strengths are 6.6\,kG for both inversions. For the 2D inversions, 287 pixels harbor fields larger than 5\,kG at $\log\tau\!=\!-0.8$ and in five of them exceeding $6.5$\,kG. It is worth noticing that the large Zeeman splitting can directly be seen in the Stokes $I$ profiles (see vertical lines in Figure~\ref{fig:inv} marking the Zeeman splitting at the magnetic field strength retrieved by the 2D inversions at $\log\tau\!=\!-0.8$).  

Table~\ref{tab:atmos} lists the average atmospheric values for both 1D and 2D inversions. The values come from the 32 pixels with $|\Vec{B}|\!>\!6.5$\,kG at $\log\tau\!=\!0$ in the 2D inversion. The parameters obtained for the pixel with the strongest field strength in the 2D inversion are given in parentheses. As the strong fields are located on both sides of the PIL, we remove the polarity information before averaging the inclination angle. Therefore, in Table~\ref{tab:atmos}, we list the inclination ranging between $0^{\circ}\leq\gamma'\leq90^{\circ} $, i.e., how inclined the field is with respect to the surface of the Sun, irrespective of its polarity. $\gamma'=90^{\circ}$ corresponds to horizontal magnetic fields (i.e., parallel to the solar surface), and the smaller $\gamma'$ is, the more vertical are the fields.

\begin{table*}[thpb]
\centering{}%
\begin{tabular}{crccccc}
\toprule[1.5pt]
Inversion & $\log\tau$ & T (K) & $|\Vec{B}|$ (kG) & $\gamma\prime$ (deg) & $v_{\rm LOS}$ (km/s)  & $\chi^2$\tabularnewline
\midrule[1.5pt]

\multirow{3}{*}{1D} & -2.3 & 4416$\pm$164 (4437) & 4.92$\pm$0.61 (5.42) & 70.5$\pm$10.4 (66.9)  & 0.8$\pm$0.8 (2.3) &\multirow{3}{*}{27.8$\pm$12.0 (36.5)} \tabularnewline
                    & -0.8 & 4808$\pm$208 (4874) & 5.82$\pm$0.64 (6.57) & 66.0$\pm$12.5 (53.3)  & 2.1$\pm$1.7 (4.9)& \tabularnewline
                    &  0.0 & 5375$\pm$361 (5114) & 6.55$\pm$0.44 (7.13)  & 58.9$\pm$13.1 (45.0) & 5.8$\pm$2.0 (6.0)& \tabularnewline
\midrule
\multirow{3}{*}{2D}  & -2.3 & 4321$\pm$245 (4395)  & 5.09$\pm$0.95 (6.65) & 68.3$\pm$12.0 (57.7)  & 1.3$\pm$1.4 (5.1)& \multirow{3}{*}{12.8$\pm$4.9 (11.5)}\tabularnewline
                     & -0.8 & 4735$\pm$288 (4909)  & 5.83$\pm$0.63 (6.57) & 62.8$\pm$12.9 (50.9)  & 2.6$\pm$2.4 (6.6)& \tabularnewline
                     &  0.0 & 5528$\pm$460 (5199)  & 6.90$\pm$0.42 (8.22) & 52.4$\pm$17.4 (28.0)  & 5.1$\pm$2.6 (3.0)&\tabularnewline
\bottomrule[1.5pt]
\end{tabular}\caption{Mean atmospheric values averaged over the 32 pixels where the 2D inversions display $|\Vec{B}|>6.5$\,kG at $\log\tau=0$. All umbral profiles were removed in order to focus only on the light bridge. 1$\sigma$ values of the 32 pixels are given as an estimation of the variation of the retrieved atmospheric parameters around the listed mean values.  The atmospheric conditions of the pixel with the largest $|\Vec{B}|$ are listed in parentheses. Note that we do not take the absolute value when averaging the $v_{\rm LOS}$ because the strongest fields were associated with downflows. $\gamma'$ ranges between 0$^{\circ}$ (vertical fields) to 90$^{\circ}$ (horizontal fields). \label{tab:atmos}}
\end{table*}

 \begin{figure}[thpb]
   \includegraphics[width=0.47\textwidth]{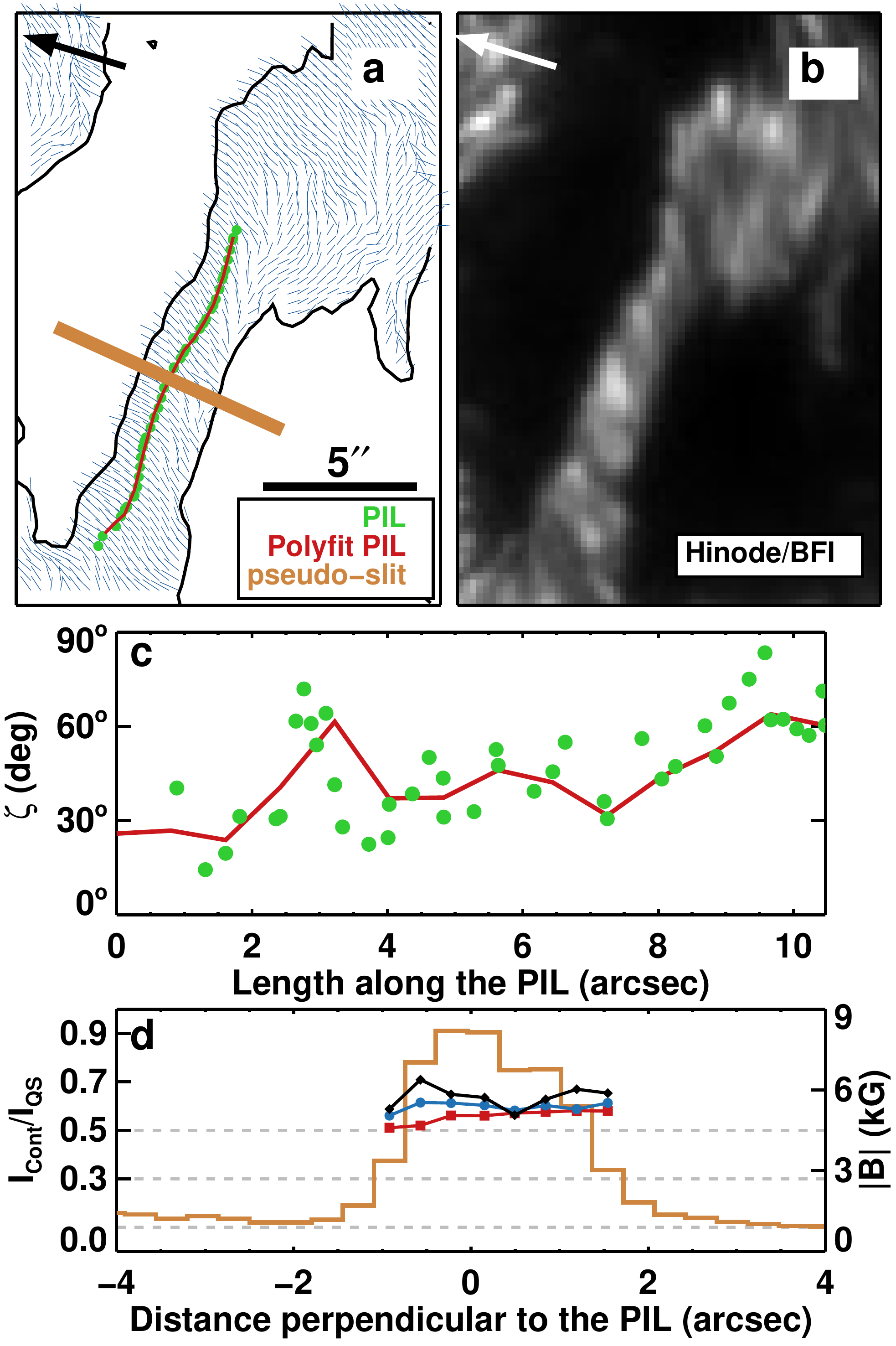}
 \caption{Azimuthal direction of the magnetic field (panel (a)) obtained by the 2D inversion at $\log\tau\!=\!0$. 
  The location of the PIL along the light bridge is indicated by the green points, and the contours are drawn at 30\%  of the quiet-Sun continuum level (panel (a)).
 The red line is the polynomial fit to the PIL. Panel (b) shows the $G$-band image taken by Hinode/BFI.  Panel (c) displays the angle, $\zeta$, between the magnetic azimuth and the ray perpendicular to the PIL at each point. That is $\zeta=0^{\circ}$ means that the field is perpendicular to the PIL, while $\zeta = 90^{\circ}$ indicates a field parallel to the PIL.  Panel (d) shows the continuum level along the pseudo-slit plotted as a brown line in panel (a), the magnetic field strength at $\log\tau\!=(0,-0.8,-2.3)$ where $I_{\text{QS}}\!>\!0.3$ (black, blue, and red lines, respectively). Gray horizontal dashed lines mark 0.1, 0.3, and 0.5 of the quiet-Sun continuum level.}\label{fig:shear}
 \end{figure}

 Table~\ref{tab:atmos} shows that the temperature in the light bridge, where the strongest fields are located, is similar to the temperature observed in regular penumbral filaments (Column (3)), which is an indication that magneto-convection is strong enough to allow hot material coming from subsurface layers to fill the light bridge.  The mean field strength at these locations is 6.9\,kG at the bottom of the photosphere and 5.1\,kG at $\log\tau\!=\!-2.3$ (Column (4)). In addition, the field is highly inclined, being somewhat  more vertical in deeper layers (Column (5)). This configuration resembles a low-lying loop-like geometry.

 Another characteristic of the strongest fields is that they are usually associated with subsonic line-of-sight velocities (Column (6)). This is in contrast to the strong-field observations at the endpoints of penumbral filaments reported by \citet{vanNoort2013A&A} and \citet{Siu-Tapia2017A&A,Siutapia2019}, which are associated with strong, supersonic downflows.

The azimuth modulo 180$^{\circ}$ inside the region of interest at $\log\tau\!=\!0$  is presented in Figure~\ref{fig:shear} (panel (a)).  The mean angle of the field with respect to the PIL is $\zeta\sim$46$^{\circ}$, exceeding $\zeta\sim$60$^{\circ}$ at the locations of the strong fields (panel (c)). The filaments visible in the intensity image are inclined at a similar angle with respect to the PIL, suggesting that the magnetic field in the light bridge is oriented along the filament direction (see Figure~\ref{fig:shear} and the animation associated with Figure~\ref{fig:video2}).

\begin{figure}[thbp]
 \begin{center}
 \includegraphics[width=.49\textwidth]{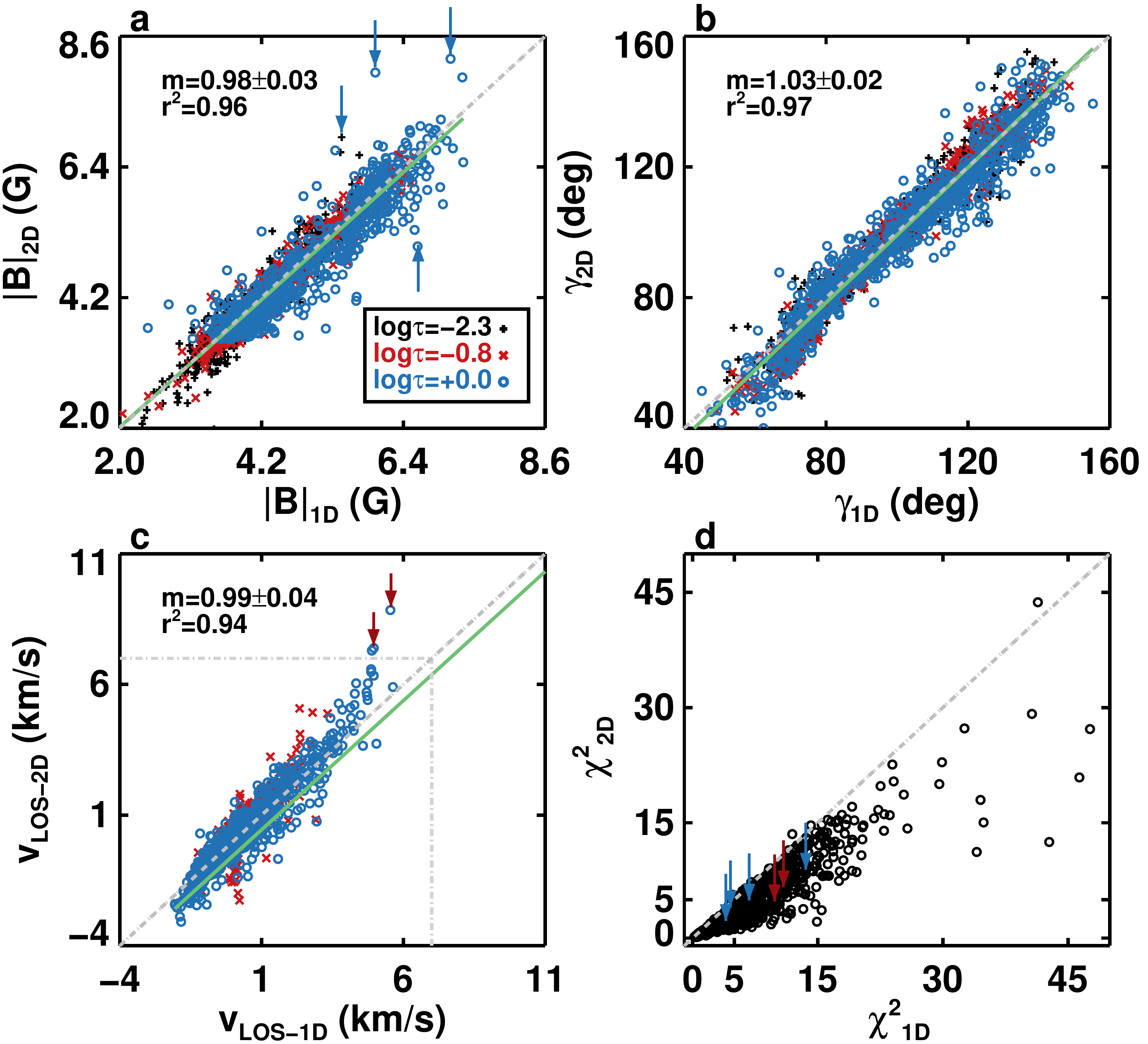}
 \caption{Scatter plots of the atmospheric parameters retrieved from 1D and 2D inversion schemes at the light bridge. Panel (a): $|\Vec{B}|$; panel (b): $\gamma$; panel (c): $v_{\rm LOS}$; panel (d): reduced $\chi^2$. Black plus signs, red crosses, and blue circles differentiate values at $\log\tau\!=\!$\,-2.3,\,-0.8, and 0.0, respectively.
 The gray dashed line is the 1:1 relationship, and the green line is the linear fit with slope $m$. The correlation coefficient $r^2$ is given in panels (a)-(c). {The blue arrows in panel (a) point to the pixels with the largest discrepancy between the 1D and the 2D approach. The red arrows in panel (c) mark 2 pixels with fast $v_{\text{LOS}}\gtrsim7$\,\kms{}. The sets of Stokes profiles of these 2 pixels and their best fits are shown in Figure~\ref{fig:special} in  Appendix \ref{sec:special}. We identified the corresponding $\chi^2$ of the pixels in panels (a) and (c) by the same colored arrows in panel (d).} }\label{fig:scatter}
 \end{center}
 \end{figure}

Since the strongest fields are returned by the 2D inversion, it is important to test how results of the 2D inversions compare with those of the generally used 1D inversions. To compare the 1D with the 2D inversion, Figure~\ref{fig:scatter} shows the scatter plots of the field strength, inclination, line-of-sight velocity, and $\chi^2$ obtained from these two types of inversions.  These scatter plots are based on the points on the light bridge with a continuum intensity $I_{\rm c}\!>\!0.3$. Choosing a different threshold does not alter the scatter plots since the boundary between the umbra and the light bridge in continuum intensity is rather sharp (see Figure~\ref{fig:shear}(d)). The scatter plots highlight the following points: 
($i$) The results from the two types of inversions agree rather well with each other. The two inversion methods do not show a systematic difference for the presented atmospheric parameters. The green lines in panels (a)-(c) show the linear fits between the results from the two inversion schemes, with correlation coefficients larger than 0.94 for the field strength and inclination and below 3\% discrepancy between the {linear fit and the 1:1 relationship.} 
The correlation for $v_{\rm LOS}$ is only slightly worse ($\chi^2$=0.94) with an 11\% slope discrepancy.
($ii$) In 55\% of the pixels inside the light bridge, the 1D inversion recovered larger values of $|\Vec{B}|$ at $\log\tau\!=\!0$. 
($iii$) The fit to the Stokes profiles using 2D inversions is by far superior compared to the 1D inversions. 99.96\% of the pixels in the $\chi^2$ scatter plot (Figure~\ref{fig:scatter}(d)) lie below the 1:1 line.

There are 11 pixels with line-of-sight downflow velocities larger than 6\,\kms{} in the 2D inversion (see Figure~\ref{fig:scatter}c). These pixels are located on the boundary between the umbra regions and the light bridge. The median field strength obtained from the 2D inversions, where the fastest flows are located, is 5.5\,kG, with a standard deviation of 2.6\,kG, and a maximum of $|\Vec{B}|=7.0$\,kG {at $\log\tau\!=\!-0.8$}. The corresponding Stokes profiles are highly complex, and the two inversion schemes return different fits. While the 2D inversion better accounts for such complex profiles by adjusting the field strength, the 1D inversions ascribe a higher $v_{\rm LOS}$ to these lines and provide a worse fit to the observations (e.g., Figure~\ref{fig:special},  row (d)).

\section{Discussion} \label{sec:discution}

The results presented in this paper confirm the presence of very strong magnetic fields in a bipolar light bridge.  \citet{Okamoto2018ApJ} reported fields of 6.2\,kG for the same region. However, the ME approach struggles in fitting the often highly complex observed Stokes profiles (see Figure 1 of \citet{Okamoto2018ApJ}) because it cannot handle height gradients that produce the asymmetries clearly visible in the observed profiles (Figures~\ref{fig:atm1d} and \ref{fig:atm2d}). The height-dependent inversions return considerably stronger fields. While the 1D inversions show these strong fields distributed over a larger area, even extending into the umbra adjacent to the light bridge, the 2D inversions concentrate the strong fields on the light bridge and in smaller regions with a maximum strength of 8.2\,kG, while providing better fits to the Stokes profiles. All along the light bridge, the magnetic field is stronger than 5\,kG for both 1D and 2D inversions. The fields are mostly horizontal (i.e. parallel to the solar surface), and their azimuth suggests that they connect the two umbrae, with an average angle of $\sim$46$^{\circ}$ {measured with respect to the normal to the PIL} (see Figure~\ref{fig:shear}). The strongest fields within the bright structure are associated with downflows \citep[as already noticed by][]{Okamoto2018ApJ} and have temperatures that are commonly found in penumbral filaments. 

\begin{figure*}[tphb!]
 \begin{center}
 \includegraphics[width=1.\textwidth]{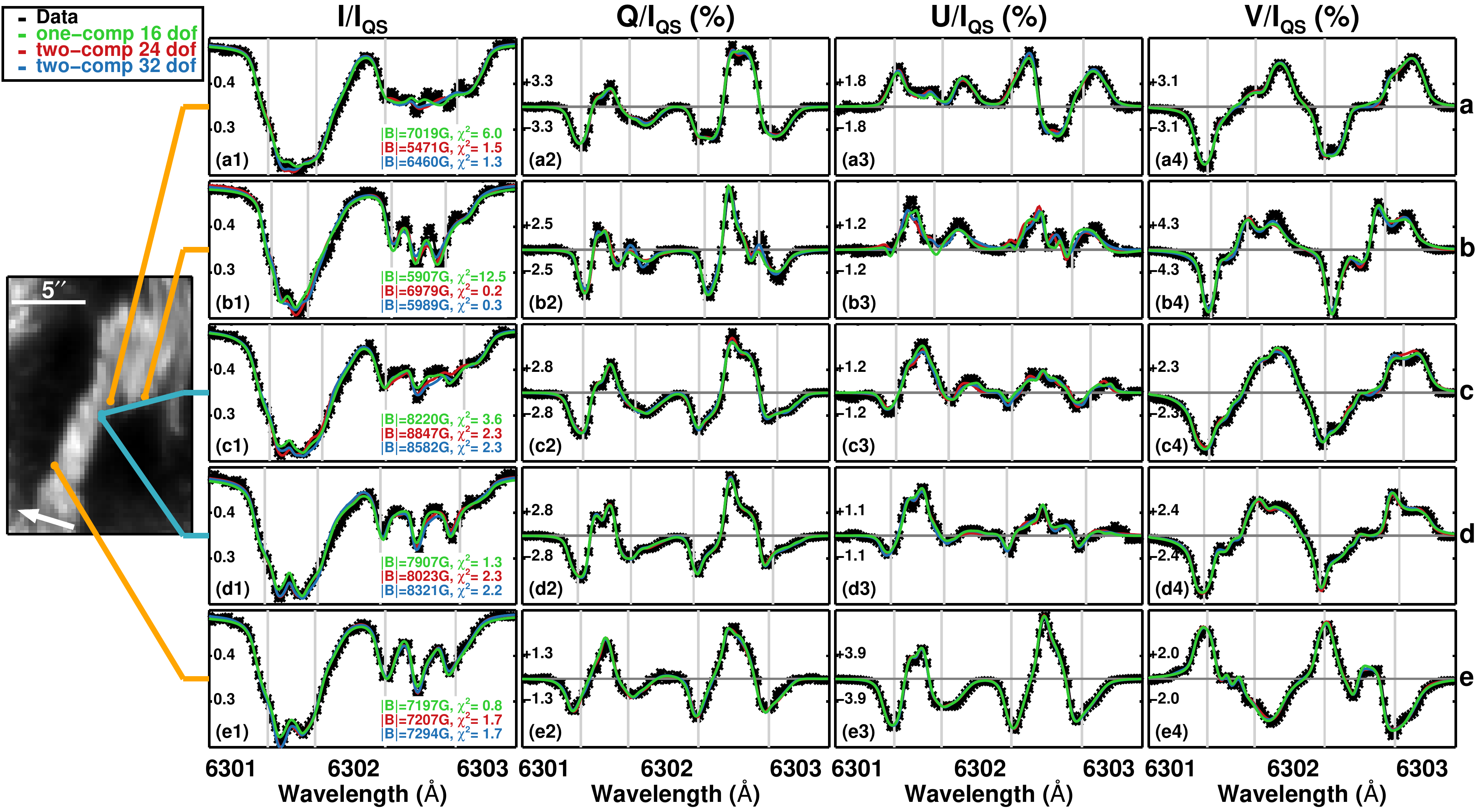}
 \caption{Same layout as Figure~\ref{fig:inv} for 5 pixels showing highly split Zeeman profiles. Black points denote the observed data. Green, red, and blue lines are the output of the 2D inversions assuming one-component atmosphere (16 dof; green), two-component atmosphere with the second component being an ME-like atmosphere (24 dof; red), and a two-component atmosphere with both having three nodes (32 dof; blue). The magnetic field strength values in the first column refer to the value of the first component at $\log\tau\!=\!0$. The difference between the three fits is minute and best  seen if the figure is magnified.
}\label{fig:2comp2d}
 \end{center}
 \end{figure*}
\begin{figure*}[tphb!]
 \begin{center}
 \includegraphics[width=1.\textwidth]{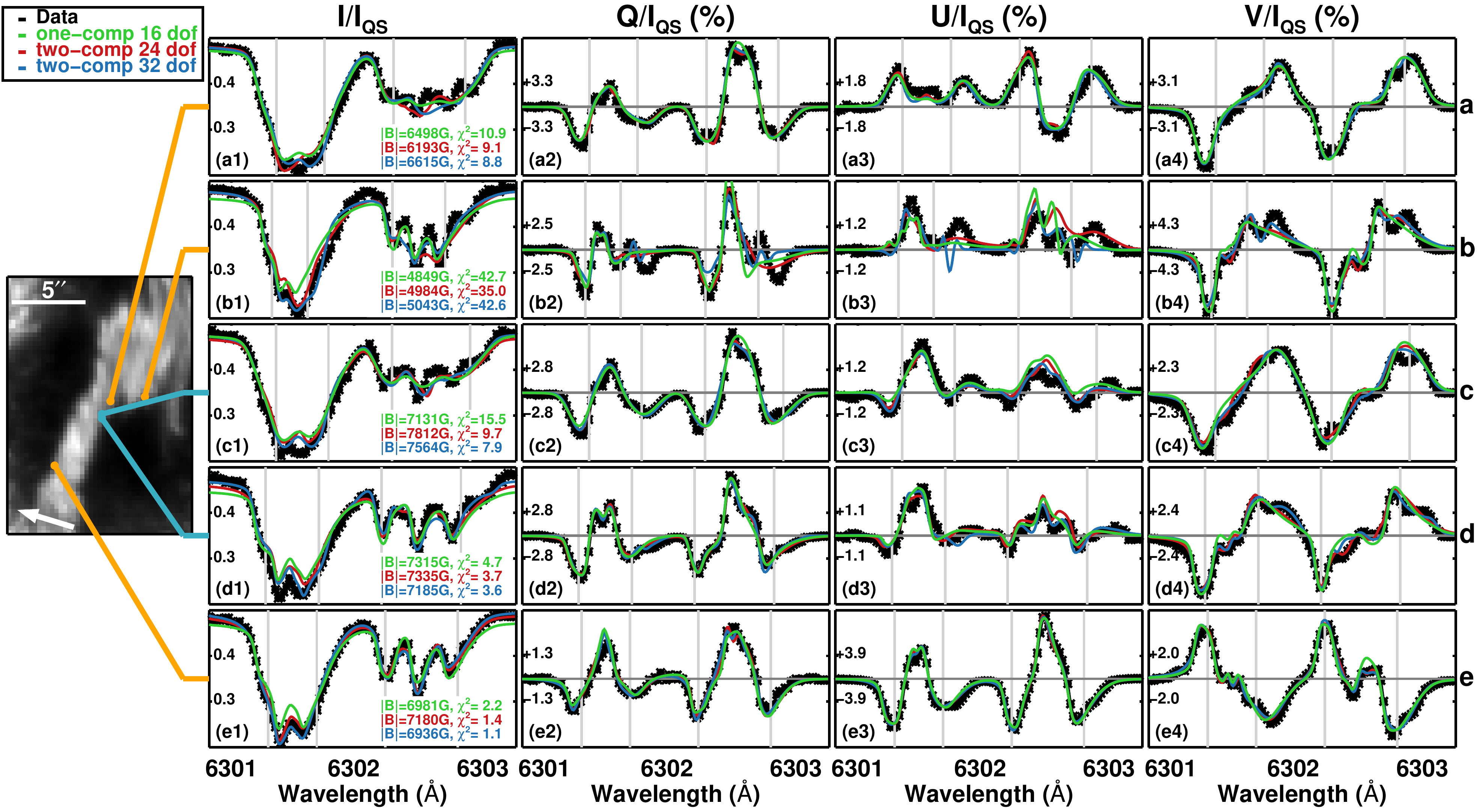}
 \caption{Same as Figure~\ref{fig:2comp2d} but for the 1D inversion. }\label{fig:2comp1d}
 \end{center}
 \end{figure*}
 
 \subsection{Need for multicomponent atmospheres?}\label{sec:2comp}
{The complexity of the Stokes profiles might be an indication of the presence of a second atmospheric component}. However, our inversion results show that the observed Stokes profiles can also be reproduced with a simple, one-component atmospheric setup that allows depth-dependent atmospheric parameters. This setup works particularly well when combined with 2D inversions, which make use of the prior knowledge about the PSF of the optical system (see Figure~\ref{fig:inv}). 
For comparison, we made experiments with two-component atmospheres for both the 1D and 2D inversion schemes (see Figures \ref{fig:2comp2d} and \ref{fig:2comp1d}).  We tested two simple atmospheric models with a second component to fit the observations. In both models, the new atmospheric component adds extra free parameters to those described in Section \ref{sec:inversions}. The models are as follows:
\begin{itemize}
    \item Model 1: Reference model (RM; Section \ref{sec:inversions}). One-component atmosphere with three nodes with 16 degrees of freedom (dof) per pixel.
    \item Model 2: RM $+$ a   second component with height-independent parameters except for the temperature. The dof are 16 for the first component, 9 for the second component (3 for $T$, and 1 each for $|\Vec{B}|$,  $\gamma$,  $\phi$,  $v_{\rm LOS}$,  $v_{\rm micro}$, and the filling factor $\alpha$). $v_{\rm micro}$ was coupled between the two components (dof per pixel: 24).
    \item Model 3: Two height-stratified components, each with three nodes for all atmospheric parameters. The dof are 16 for both the first and the second component ($T$, $|\Vec{B}|$, $\gamma$, $\phi$, and $v_{\rm LOS}$ vary along the line of sight, each contributing with three free parameters, as well as the  filling factor $\alpha$). $v_{\rm micro}$ was coupled between the two components (dof per pixel: 32).
\end{itemize}

Fitting a second component (Model 3) to the complex profiles in the light bridge requires a careful selection of the initial conditions for the inversion to ensure convergence to the global minimum, especially when one of the components is dominant. We tested different limits for the free parameters velocity and field strength. We also measured the velocity of the possible second component ($v'_{r\rm ed}\approx18$\,\kms{}), and we used it as the initial condition of the more rapidly downflowing component. We also tried bounding the second component to values $\pm5$\,\kms{} around the $v'_{\rm red}$ value. In other experiments, we varied the initial values for both components of field strength and velocity. We also tried to initialize the inversion with a second component of opposite polarity, or to impose the filling factor for the second component. None of our tests produced results that deviated significantly from the values of the strong fields that we report below. 

Figures \ref{fig:2comp2d} and \ref{fig:2comp1d} show examples from 5 pixels with strong fields and with a possible second component. The green, red, and blue lines show the best fit using models 1 to 3, respectively. The corresponding values for the reduced $\chi^2$, $|\vec{B}|$, $\gamma$, $v_{\text{LOS}}$ and the filling factor $\alpha$ for the three models are listed in Table~\ref{tab:atmos2comp} in Appendix \ref{sec:experiments}. Our experiments provide clear evidence for the existence of strong fields irrespective of the model: the observations are well fitted, and the large fields ($>$6\,kG)  appeared in both 1D and 2D inversions, even when applying a two-component atmosphere model. 

In addition, our experiments  did not find that a second component significantly increased the quality of the fit to the observed Stokes profiles. The filling factors for the second component are usually small ($\lesssim$20\%), while the reduced $\chi^2$ does not improve with increasing number of free parameters, in particular for the 2D inversion. There is some improvement for the 1D inversions, which is expected (see below). 

The need for a second component within the resolution element is obviously only necessary to account for the photons originating from the strong-field regions in the light bridge that are distributed over a larger area by the telescope PSF or to mimic the effects of vertical gradients in the atmosphere. Figures \ref{fig:atm1d} and \ref{fig:atm2d} show the existence of such gradients, which cannot be reproduced by the height-independent ME-type atmospheres \citep[such as the inversion scheme applied by][]{Okamoto2018ApJ}. 
For these reasons, we can safely state that a second atmospheric component within the resolution element is not required to reproduce the complex observed Stokes profiles.

\subsection{Temporal evolution of the light bridge} \label{sec:temporal}

The AR 11967 appeared on the east limb on 2014 January 28. On the 29th, two opposite-polarity umbrae merged and formed a large, complex sunspot. The two opposite-polarity umbrae coexisted over the whole time while transiting the solar disk.  The region between the two umbrae was occupied by the light bridge harboring the strong fields. This light bridge remained there with varying thickness and intensity over the whole time the AR crossed the solar disk. The negative-polarity ($\gamma <0$) umbra adjacent and east of the light bridge in Figure~\ref{fig:ar} moved in a northwesterly direction, while the umbra adjacent and west of the light bridge moved in a southeasterly direction.
The light bridge appears exactly at the PIL between both opposite-polarity umbrae (see Figure~\ref{fig:video1}). As a consequence, this light bridge is not of the classical type that usually appears in a decaying umbra with a single polarity. 

To understand the existence of the strong fields in the light bridge, measured around 19:00\,UT on 2014 February 4, we study its temporal evolution.  There were two flares before the observation of the strong fields at 16:02\,UT and 18:49\,UT and another at 19:41\,UT after the observations ended (see Table~\ref{tab:localflares}). SDO/AIA images suggest that these flares were associated with the light bridge. The photospheric images taken by HMI and Hinode/BFI reveal the formation of a very long filament connecting the light bridge on one end to a pore moving away from the spot on the east side of the AR (see Figure~\ref{fig:video3}). This long filament breaks down around 16:00\,UT, the time of the maximum of Flare 1. The breakdown seems to be connected with brightenings in the \ion{Ca}{2} images. Then, for about an hour, the light bridge separating the two opposite-polarity umbrae almost vanished. After this, a \textit{new} light bridge formed at the PIL with a filamentary structure inclined by $\sim$46$^{\circ}$ with respect to the PIL. The  light bridge broadened until it reached a width of 3 Mm at 19:11\,UT. 
Flare 2 is observed between 18:36 and 18:54\,UT before the Hinode/SP observations at the light bridge that took place between 19:10 and 19:12\,UT. Finally, 15 minutes later, GOES reported the start of a C5.4 class flare at 19:27\,UT.
Another possibility for the appearance of the light bridge would be that the two umbrae move away from each other. However, with this process it would be more difficult to explain the strong fields on the light bridge. Further information about all flares produced by the AR 11967 is summarized in Table~\ref{tab:flares}.

It is unclear whether the disappearance of the light bridge is caused by Flare 1. It should be noted that the maximum field strength in this light bridge might have been reached between the two Hinode/SP scans at 15:42\,UT and 19:00\,UT, but there are no Hinode/SP observations available to prove it.

 \begin{table}[htbp]
\centering
\begin{tabular}{ccccc}
\toprule[1.5pt]
&GOES  & Begin & Maximum & End   \tabularnewline
&Class & (UT)  & (UT)    & (UT)  \tabularnewline
\midrule[1.5pt]
Flare 1&M1.5  & 15:25 & 16:02   & 16:40 \tabularnewline
Flare 2&C4.7  & 18:36 & 18:49   & 18:54 \tabularnewline
Flare 3&C5.4  & 19:27 & 19:41   & 19:54 \tabularnewline
\bottomrule[1.5pt]
\end{tabular}
\caption{Flares at the light bridge from 15:00\,UT to 20:00\,UT on 2014 February 4. \label{tab:localflares}}
\end{table}

\subsection{Possible mechanisms to amplify the magnetic field}

The main aim of this paper is the reliable determination of the strong magnetic fields in the light bridge. Nonetheless, in this section we briefly sketch some possible mechanisms for the amplification of the magnetic field to such high values. More insight into the mechanisms can be gained by performing a statistical study of different ARs with a similar, opposite-polarity configuration of umbrae. We already identified such regions and will report the results in a follow-up paper.

\subsubsection*{Looking deeper in the atmosphere}
The strong fields are located in a light bridge and not in the umbra. The temperature inside the structure is similar to penumbral filaments; therefore, the opacity is expected to be higher than in the umbra owing to the higher temperature (for a fixed field strength). The very strong fields, however, imply a strong evacuation of the gas due to horizontal force balance, which in turn lowers the opacity, so that where the strong fields are measured we are seeing deeper layers in which fields are typically stronger. Downflows also may enhance the evacuation, although for subsonic downflows the effect will likely be limited. It is worth mentioning that assuming that the flows inside the light bridge are field aligned ($v_{\text{f-a}}\!\!=\!\!v_{\text{LOS}}/\cos\gamma$), the velocities in some pixels reach values close to or even somewhat above  the sound speed, which is about $c_s\approx6.9$\,\kms{} at $\log\tau\!=\!-0.9$. For the 8.2\,kG pixel, the field-aligned velocity reaches its maximum of $\sim$8.9\,\kms{} at $\log\tau\!=\!-0.9$. However, in most of the pixels with strong fields the field-aligned velocities are subsonic at all heights. In particular, the field-aligned velocities are far from the 22\,\kms{} previously reported for strong-field regions at the outer boundary of the penumbra of a sunspot  \citep{vanNoort2013A&A}, or at the tails of counter-Evershed flows \citep{Siu-Tapia2017A&A}. Consequently, the evacuation and reduced opacity helps to explain the very strong fields but is on its own not sufficient.

\subsubsection*{Amplification of the field caused by downflows at a magnetic barrier}

Downflows have been proposed to enhance magnetic field strengths by the convective collapse instability \citep{Parker1978ApJ,Spruit1979SoPh,Grossmann-Doerth1998A&A}. The $>$7\,kG strong fields reported by \citet{vanNoort2013A&A} have large downflow velocities of up to 22 \kms{} associated with them \citep[see][]{EstebanPozuelo2016ApJ}. \citet{Siutapia2019} reported strong fields in a region occupied by counter-Evershed flows. Those strong fields were also associated with fast downflow velocities, where the umbra acts as a magnetic barrier, compressing the channels along which the downflows occur. The large downflow velocities observed in both studies are also observed in MHD simulations, which means that such solutions can be explained using the assumptions underlying the simulations. 
Despite many similarities to these works, our situation is different in one aspect:
we observed that 85\% of the strong fields exceeding 6.5\,kG are associated with subsonic downflows, with velocities of 5\,\kms{} on average and 3\,\kms{} for the strongest field. 
Nevertheless, the process of amplifying the field could be the same: magnetized plasma hits a barrier (i.e., the umbra), where it is forced to flow down at the endpoints of these filaments associated with strong and more vertical magnetic fields at deeper layers.

\citet{Okamoto2018ApJ} suggested that the magnetic field lines at the light bridge are subducted by the Evershed flow. The strongest fields, however, appear in the filaments carrying a flow toward the umbra, opposite to the normal Evershed flow direction \citep[as pointed out by the referee; see footnote 6 in][]{Okamoto2018ApJ}. 
However, in the present configuration of a penumbra existing between two umbrae it is not even clear what the normal Evershed flow direction should be.
Nevertheless, it can be stated that the strong magnetic field strength within the bipolar light bridge of the AR 11967 is associated with downflows, avoiding the association with Evershed flows. 

\subsubsection*{Magnetic flux emergence}

The bipolar light bridge harboring strong fields can also be an example of an emerging photospheric flux rope similar to the one reported by \citet{Guglielmino2017ApJ,Guglielmino2019ApJ} and \citet{Bharti2017A&A}. A flux rope is inherently associated with strong fields buoyantly rising from subsurface layers. From the temporal evolution observed by HMI and Hinode/BFI, the light bridge seems to be twisted (see the online animations) and increases in width with time. In addition, the negative-positive Doppler velocity pattern obtained from the inversions  tends to agree with this picture \cite[]{Lites2010ApJ}.

The special configuration  of two adjacent opposite-polarity umbrae offers an alternative flux emergence scenario: 
the previously field-free gas, trapped between the two opposite-polarity umbrae in subsurface layers, advects the magnetic field of the adjacent flux tubes connected to the two umbrae. This advection occurs during the motion within the elongated convection cell forming the filamentary channels observed on the light bridge. 

\subsubsection*{Shear-induced field amplification}

The light bridge consists of filaments with a tilt angle of 40$^{\circ}$ to 60$^{\circ}$ with respect to the PIL. This tilt could be the result of a shear, caused by the motion of the two umbrae relative to each other, which would have amplified the field through induction \citep[see e.g.,][]{Toriumi2019ApJL,Anfinogentov2019ApJ}.

\section{Summary and Conclusion}\label{sec:conclusion}

The thermodynamic and magnetic field properties of the observed light bridge separating two umbrae of opposite polarity can be summarized as follows:
\begin{itemize}
    \item We find fields larger than 5\,kG at all three optical depth node points used during the inversions. The area covered by fields $>$5\,kG is 32.7\,arcsec$^2$ at $\log\tau\!=\!0$, and 26.6\,arcsec$^2$ at $\log\tau\!=\!-0.8$. Fields larger than 6.5\,kG are observed in 32 pixels distributed over two contiguous regions. In nine of these pixels the field strength exceeds $7$\,kG. We report evidence of a record-high field strength of 8.2\,kG at $\tau\!=\!1$ in one pixel.
    \item Coupled 2D inversions \citep{vanNoort2012A&A} provide better fits than 1D inversions but give on average similar $|\vec{B}|$ values. However, for the strongest fields, 2D inversions give larger field strength. 
    \item The best fit to the observed Stokes profiles does not require a second atmospheric component. However, in all tested inversion setups (height-dependent one-component and two-component atmospheres of different complexity), the strong magnetic fields are reproduced.
    \item The fields are mainly horizontal (see Figures~\ref{fig:atm1d}and \ref{fig:atm2d}), and the fields are observed to be more vertical at deeper layers, independent of their polarity (see Table~\ref{tab:atmos}). 
     \item The angle of the magnetic field with respect to the PIL is on average $\sim$46$^{\circ}$.
     At the places with the strongest magnetic fields, this angle is $\sim$60$^{\circ}$ {(see Figure~\ref{fig:shear}).}
     \item  The $v_{\rm LOS}$ is higher in deeper layers. {The positive and negative Doppler velocities} are approximately symmetric and separated by the PIL. The strongest fields are associated with downflows. A total of 32 pixels with fields larger than 6.5\,kG have subsonic line-of-sight velocities.
    \item The temperature stratification inside the light bridge is similar to the temperature stratification inside penumbral filaments.  This indicates that the filaments in the light bridge have a similar origin to penumbral filaments, where magneto-convection is supplying the filaments with hot material from deeper layers. 

\end{itemize}

The sum of our observational findings suggests that the light bridge is composed of a twisting and likely emerging flux rope, still largely buried under the surface at the time of the analyzed observations. However, this and other potential interpretations will be tested on a larger data set in an upcoming publication.

The largest field observed in this analyzed light bridge has a magnitude of 8.2\,kG ($=$0.82\,tesla). A systematic study of similar configurations of ARs, where a light bridge separates two umbrae of opposite polarities, could provide insight into how common such extremely strong magnetic fields are and if even stronger fields exist on the Sun. An important step toward the discovery of even stronger fields would be high spatial resolution observations of such a light bridge using the infrared Fe lines located at 1.56\,$\mu$m, sampling the deepest observable layers of the photosphere \cite[e.g.,][]{Solanki1992A&A,Milic2019}.

\acknowledgements 
We thank Maarit K\"apyl\"a and the SOLSTAR group, for discussions about possible field amplification mechanisms, and J.\,Okamoto whose seminar talk at MPS inspired this work. {We thank the referee for careful reading of the manuscript and insightful comments}. J.S.C.D was funded by the Deutscher Akademischer Austauschdienst (DAAD) and the International Max Planck Research School (IMPRS) for Solar System Science at the University of G\"ottingen. This project has received funding from the European Research Council (ERC) under the European Union’s Horizon 2020 research and innovation program (grant agreement No. 695075) and has been supported by the BK21 plus program through the National Research Foundation (NRF) funded by the Ministry of Education of Korea. Hinode is a Japanese mission developed and launched by ISAS/JAXA, with NAOJ as domestic partner and NASA and UKSA as international partners. It is operated by these agencies in cooperation with ESA and NSC (Norway). {We thank the SDO/HMI team members for the data. SDO is a mission of the NASA's Living With a Star program.}

\begin{appendix}

\section{Node position} \label{sec:nodesposition}

The choice of the location of the nodes  can affect the fits of the profiles, the results, and even in some cases the interpretation. 
Only results that are robust against the exact placement of the nodes can be considered to be reliable. 
For the values reported in this work, the locations of the nodes were optimized to obtain the global minimum for the entire map. To test the reliability of the reported strong fields, we performed a set of experiments by placing one of the nodes at a different optical depth using the 2D inversion scheme.  Table~\ref{tab:nodes} presents the field strength retrieved at $\log\tau\!=\!(0,-0.8,-2.3)$, the reduced $\chi^2$, and the mean reduced $\chi^2$ over the area influenced by the PSF. For the sake of the example, we chose the same pixel reported harboring 8.2\,kG magnetic field strength (Table~\ref{tab:nodes}(a)).

\begin{table*}[htbp]
    \centering
\begin{tabular}{crccccc}
\toprule[1.5pt]
 \multirow{ 2}{*}{{Test}} & \multicolumn{1}{c}{{Node Location}}  & {$|\Vec{B}|_{\log\tau=0}$} &{$|\Vec{B}|_{\log\tau=-0.8}$}&{$|\Vec{B}|_{\log\tau=-2.3}$} & \multirow{ 2}{*}{{$\chi^2$}} & \multirow{ 2}{*}{{$\overline{\chi}^2$}}\\ 
& \multicolumn{1}{c}{{($\log\tau$)}}& {(kG)}& {(kG)}& {(kG)} & &\\
\midrule[1.5pt]
a&{(0.0, -0.8, -2.3)} & {8.22} & {6.57}  &{6.65}&{3.6}& {1.8} \\ 
\cmidrule(l{0.2cm}){0-1}
\multirow{ 3}{*}{b}&{(+0.05, -0.8, -2.3)} & {8.15} & {6.66} &{5.88}&{3.5}&{1.8}\\
&{(+0.1, -0.8, -2.3)} & {8.23} & {6.79}&{5.79}&{3.7} &{1.9}\\ 
&{(+0.5, -0.8, -2.3)} & {8.09} & {6.97} &{6.35}&{4.0}&{1.9}\\
\cmidrule(l{0.2cm}){0-1}
\multirow{ 2}{*}{c}&{(0.0, -0.6, -2.3)} & {7.84} & {6.50}&{6.23}&{4.2}&{2.0} \\
&{(0.0, -1.0, -2.3)} & {8.23} & {7.05}&{5.81}&{3.7}&{1.8}\\
\cmidrule(l{.2cm}){0-1}
\multirow{ 2}{*}{d}&{(0.0, -0.8, -2.0)} & {7.94} & {6.27}&{6.58}&{3.5}&{1.8}\\
&{(0.0, -0.8, -2.5)} & {7.54} & {6.77}&{6.07}&{4.2}&{2.0}\\
\cmidrule(l{.2cm}){0-1}
e&{(0.0, -0.8, -2.3)}  & {7.93} & {6.97}  & {6.35}&{3.1} &{1.8} \\ 
\cmidrule(l{1.8cm}r{.5cm}){2-5}
\multicolumn{2}{r}{{Median}} & {8.09} & {6.77} & {6.23} & & \\
\multicolumn{2}{r}{{Mean}} & {8.02} & {6.72} & {6.19} & & \\
\multicolumn{2}{r}{{$\sigma$}} & {0.23} & {0.26} & {0.32} & & \\
\bottomrule[1.5pt]
\end{tabular} 
    \caption{{Variation of the field strength at the strongest field pixel depending on the location of the node position. (a) Nodes setting chosen in the paper, (b) bottom node below the $\log\tau\!=\!0$, (c) and (d) are the cases where the location of the middle/top node was slightly changed. Test (e) shows the experiment with only two nodes in the line-of-sight velocity. The goodness-of-fitness is given by the reduced $\chi^2$, and $\overline{\chi}^2$ is the mean reduced $\chi^2$ over the pixels influenced by the PSF. }
    \label{tab:nodes}}
\end{table*}

In test (b), we placed the bottom node below $\log\tau\!=\!0$ at the locations of  $\log\tau\!=\!(+0.05,+0.1,+0.5)$.  
As it can be seen from Table~\ref{tab:nodes}, the results do not depend strongly on the location of the bottom node. The standard deviation of the field strength between these experiments is 70, 160 and 300\,G for the bottom, middle, and top nodes, respectively. 
In the case of the middle, and top nodes, we selected two places above and below the node position used in the paper (Table~\ref{tab:nodes}, tests (b)/(c)). The magnetic field strength in these cases still shows the mean value of $7.89\pm0.28$\,kG at $\log\tau\!=\!0$.

 \begin{figure*}[tphb]
 \begin{center}
    \includegraphics[width=0.65\textwidth]{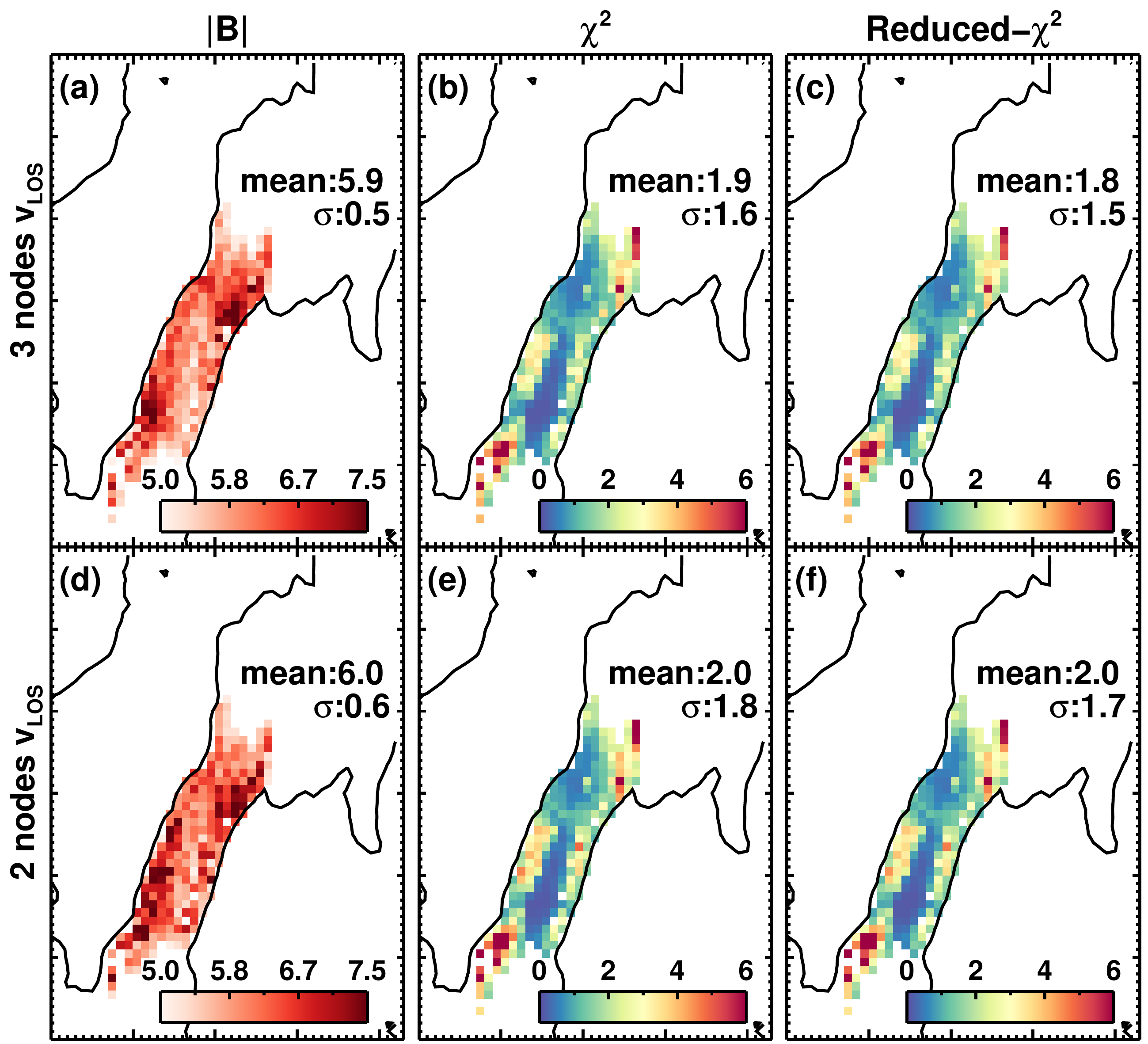}
    \caption{Comparison between the three-node inversion (top) and the two-node inversion (bottom) for the $v_{\rm LOS}$. Field strength, $\chi^2$, and the reduced $\chi^2$ are shown from left to right.}
\label{fig:chi2}
 \end{center}
 \end{figure*}

Additionally, we performed inversions with two nodes in the velocity. Figure~\ref{fig:chi2} shows the comparison between the three-node (top) and two-node (bottom) inversions. For the case of the 8.2 kG pixel, the reduced $\chi^2$ is smaller compared to the $\chi^2=3.6$ when having three nodes. However, for other pixels within the light bridge, the magnetic field and the reduced $\chi^2$ values are larger (see Figures~\ref{fig:chi2} (d)-(f)). The mean $\chi^2$ value increases from 1.7 to 1.9, with a larger standard deviation changing from 1.4 to 1.6. For the strongest field pixel, the magnetic field decreased by 3.5\% at $\log\tau\!=\!0$. On the other hand, the mean field strength over the entire light bridge increased by 12.5\% (see left panels in Figure~\ref{fig:chi2}). This increment in the magnetic field can be easily explained: to fit the asymmetries in the wings of the spectral line, the code assigns larger magnetic field strength to compensate the missing information carried by the line-of-sight velocity.

\begin{figure*}[htpb]
 \begin{center}
 \includegraphics[width=.99\textwidth]{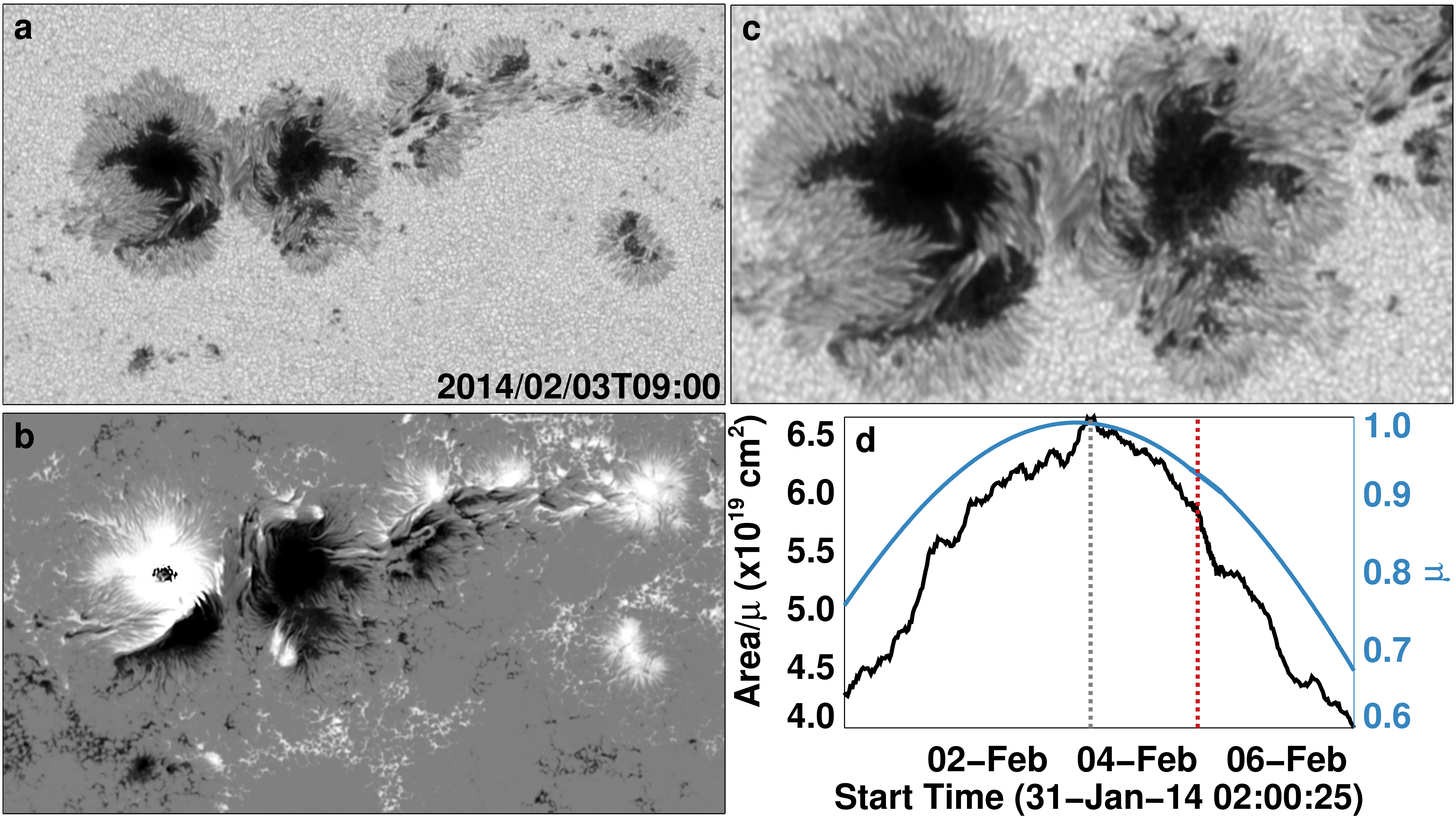}
 \caption{Snapshot of online animation 1. The temporal evolution of the AR 11967 crossing the solar disk between $\pm45^{\circ}$ with a cadence of 30 minutes as observed by SDO/HMI. The snapshot was taken at the time when the AR 11967 reaches its maximum covered area. Left panels show (a)  the continuum intensity and (b) the magnetogram  clipped at $\pm1.5$\,kG. Panel (c) shows a zoom in the region. Panel (d) presents the area covered (black line) by the AR 11967 corrected by foreshortening. The blue line is the cosine of the heliocentric angle $\mu$. Vertical dashed lines exemplify the time of the frame (gray) and the time Hinode/SP observation (red). The video begins on 2014 January 31 at 02:00\,UT and ends 21\,s later, displaying the evolved AR 7 days later on 2014 February 6 at 21:30\,UT.
}\label{fig:video1}
 \end{center}
 \end{figure*}

The mean values of the magnetic field between all the experiments are  $|\vec{B}|(\log\tau\!=\!0)=(8.09\pm0.23)$\,kG, $|\vec{B}|(\log\tau\!=\!-0.8)=(6.77\pm0.26)$\,kG, and  $|\vec{B}|(\log\tau\!=\!-2.3)=(6.23\pm0.32)$\,kG. This clearly supports the strong-field character of the light bridge. Therefore, the location of the bottom node seems to not affect strongly the reported results.

\begin{figure}[htbp]
 \begin{center}
 \includegraphics[width=0.47\textwidth]{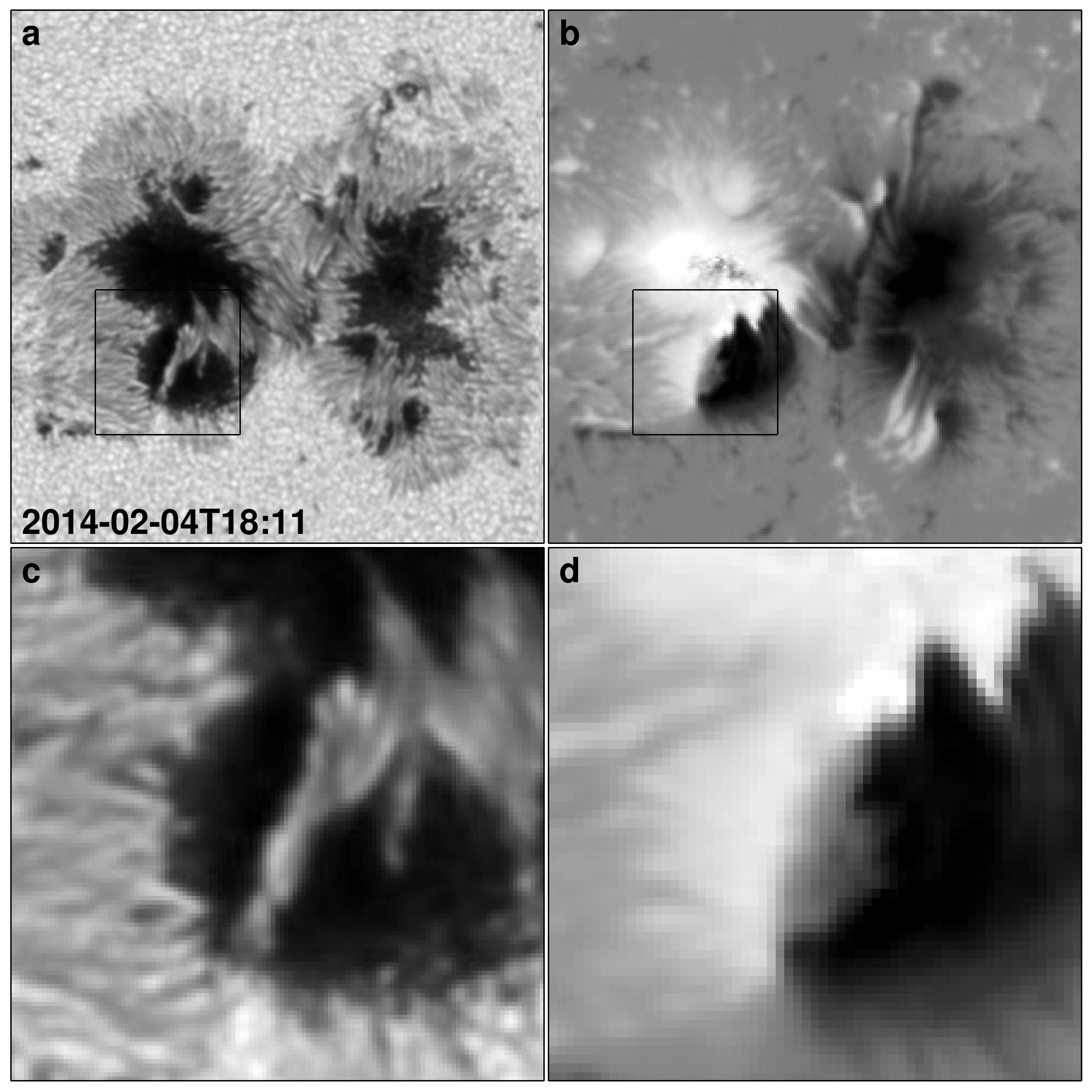}
 \caption{Snapshot of online animation 2. Temporal evolution of the AR 11967 during 10 hr around the observation of strong fields with a cadence of 45 seconds as observed by SDO/HMI.  Left panels show the continuum intensity and right panels show the magnetograms clipped at clipped at $\pm2$\,kG. Bottom panels  are the scene within the black squares displayed on the top panels. The video begins on 2014 February 4 at 17:00\,UT and ends 33 seconds later displaying the last images taken on 2014 February 5 at 02:58\,UT.
 }\label{fig:video2}
 \end{center}
 \end{figure}
 
\newpage
\section{Online material}\label{sec:videos}

\begin{figure}[htbp]
 \begin{center}
 \includegraphics[width=0.47\textwidth]{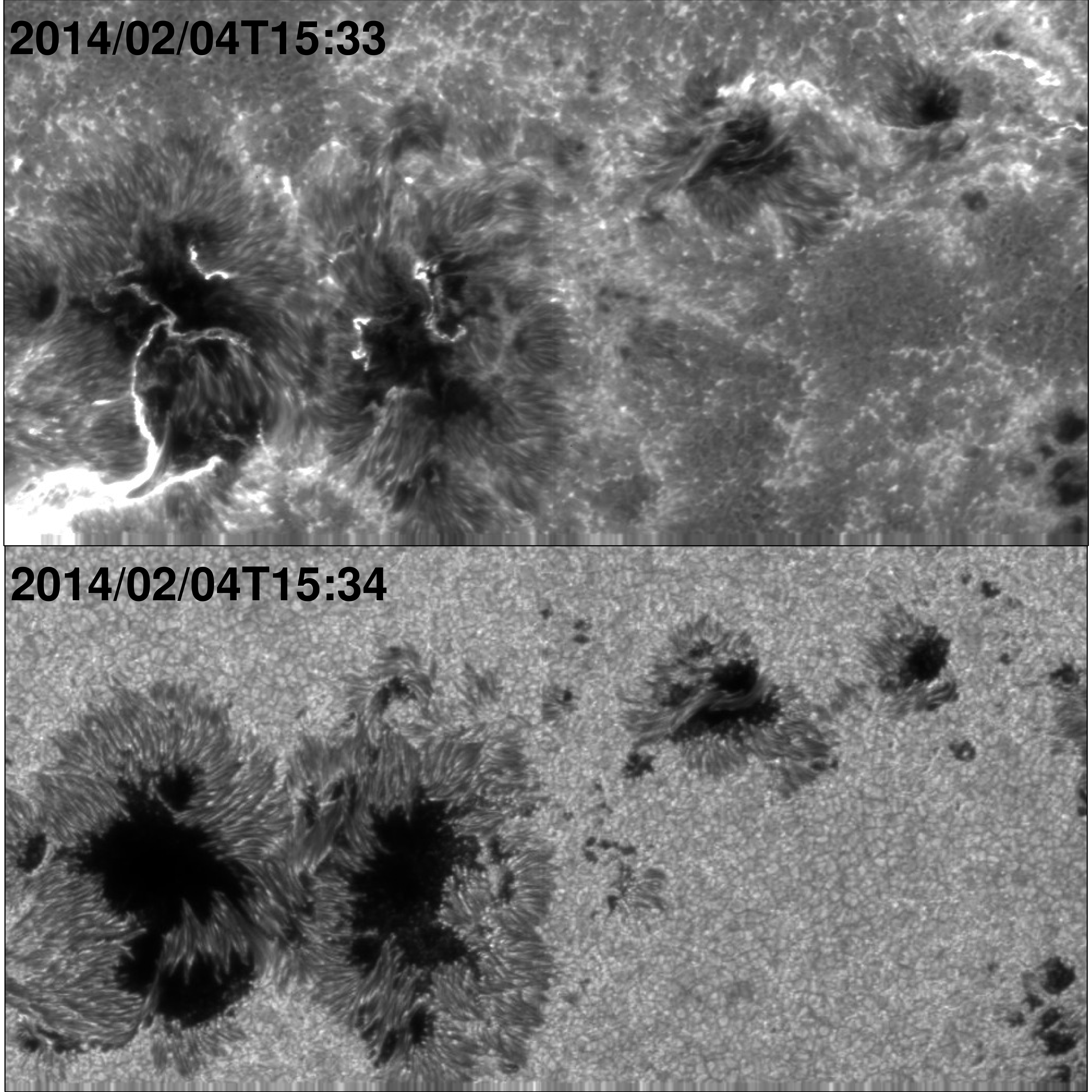}
 \caption{Snapshot of online animation 3. Hinode/BFI observations during 26\,hr starting on February 4 at 00:00\,UT. Top panel: filtergram images of the \ion{Ca}{2}\,h at $3968.5$\,\AA{} with 1-minute cadence. Bottom panel: $G$-band filtegrams around $3883.5$\,\AA{} with 10-minute cadence. The video begins on 2014 February 4, displaying the \ion{Ca}{2} image taken at 00:00 UT and the first $G$-band image image taken at 00:39\,UT. While the \ion{Ca}{2} images run continuously, the $G$-band images are updated every 10 minutes. The video ends the next day at 02:00\,UT in the top panel and 5 minutes before in the bottom one. The duration of the video is 67 seconds. 
 }\label{fig:video3}
 \end{center}
 \end{figure}

The three videos are provided as online material and display the evolution of the AR as it was observed by SDO/HMI and Hinode/BFI. Figure~\ref{fig:video1}, \ref{fig:video2} and \ref{fig:video3} present examples of snapshots of the videos and describe their layouts.

 \section{Some noteworthy Stokes profiles inside the region of interest}\label{sec:special}

  \begin{figure*}[tphb!]
 \begin{center}
 \includegraphics[width=.95\textwidth]{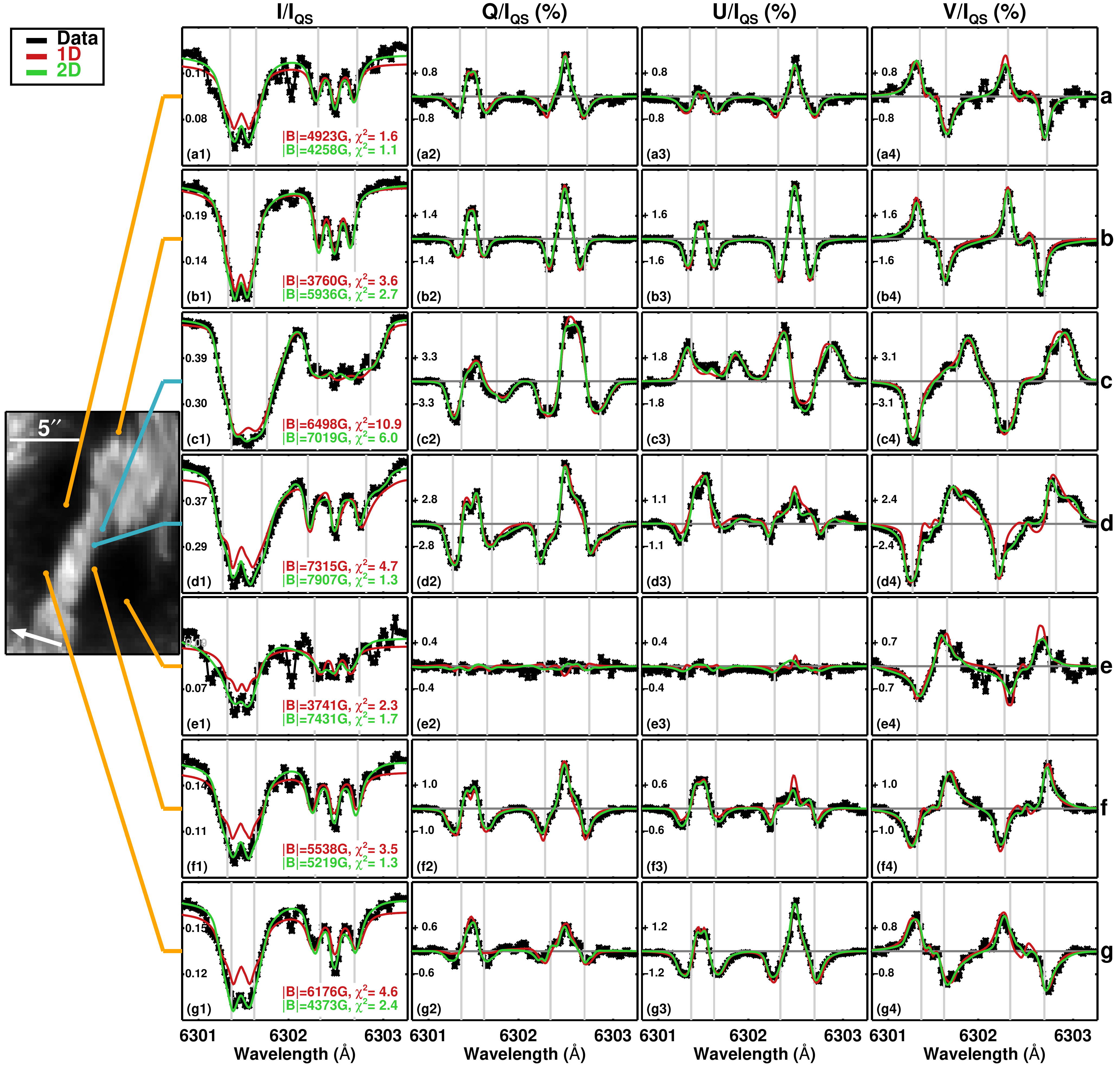}
 \caption{Observed Stokes profiles and the fits from the 1D and 2D inversion schemes. {Values of the field strength at $\tau\!=\!1$ are given in the first column.}  Same layout as Figure~\ref{fig:inv}. }\label{fig:special}
 \end{center}
 \end{figure*}

Examples of special regions at/next to the light bridge are presented in Figure~\ref{fig:special}. Rows (b), (d), (f), and (g) show the outliers pointed out in Figure~\ref{fig:scatter}, where the $|\vec{B}|$ from 1D and 2D inversions presented differences larger than 1\,kG at $\log\tau\!=\!0$, or at $\log\tau\!=\!-0.8$ (row (f)), or difference in $v_{\text{LOS}}$ (row (d)). Rows (c) and (d) show profiles that result in very strong magnetic fields associated with  superfast velocities $\gtrsim$6.9\,\kms{}  (the sound speed in the penumbra ranges from $\sim$6 to $\sim$8\,\kms{}). 

For completeness in rows (a) and (e), we display common dark umbra profiles that are clearly contaminated with molecular lines, excluded in our analysis. Profiles from rows (b), (f), and (g) show similar contamination effects, although not as strong as in row (e), since the temperature in those regions is still high enough to prevent molecules from fully dominating these profiles.

 \section{Flare production of the AR 11967}\label{sec:flares}
It is worth noting that such a large AR, harboring fields $>$5\,kG, produced only small flares. We checked GOES 1-8\,\AA{} for those days when the region crossed the solar disk. Table ~\ref{tab:flares}
summarizes the total number of C- and M-class flares per day
that were hosted by the AR 11967 from 2014 January 26 to
February 9. Row (4) marks the largest flare of the day. The largest flare was an M6-class flare that occurred 5 days before the 8\,kG fields were observed (see Table~\ref{tab:flares}). 
The low activity of AR 11967 contrasts with that of AR12673 presented by \cite{Wang2018RNAAS}. The AR 12673 presented similarly strong fields {\citep[see, e.g.,][]{Anfinogentov2019ApJ}}, but produced multiple X-class flares \citep[e.g.,][]{Verma2018A&A,Romano2019SoPh} and a coronal mass ejection \citep{Veronig2018ApJ}, including the largest flare of the 24th solar cycle \citep[e.g.,][]{Hou2018A&A,Jiang2018ApJ}.  We confirmed the existence of the strong fields by applying both 1D and 2D inversions to the Hinode/SP data of AR 12673, but this discussion goes beyond the scope of the present paper \citep[see also review by][]{Toriumi2019LRSP}.

\begin{table}[htbp]
    \centering
\begin{tabular}{lccc}
\toprule[1.5pt]
\multirow{2}{*}{Date} & \multirow{2}{*}{C-class} & \multirow{2}{*}{M-class} & Largest flare \tabularnewline
 &  &  & of the day\tabularnewline
\midrule[1.5pt]
01/26 & 1 & 0 & C1.5\tabularnewline
01/27 & 3 & 3 & M4.9\tabularnewline
01/28 & 5 & 6 & M4.9\tabularnewline
01/29 & 4 & 0 & C7.3\tabularnewline
01/30 & 6 & 3 & M6.6\tabularnewline
01/31 & 5 & 0 & C6.3\tabularnewline
02/01 & 6 & 2 & M3.0\tabularnewline
02/02 & 2 & 5 & M4.4\tabularnewline
02/03 & 9 & 0 & C9.0\tabularnewline
02/04$^{a}$ & 8 & 3 & M5.2\tabularnewline
02/05 & 5 & 1 & M1.3\tabularnewline
02/06 & 3 & 1 & M1.5\tabularnewline
02/07 & 3 & 1 & M2.0\tabularnewline
02/08 & 6 & 0 & C8.6\tabularnewline
02/09 & 2 & 0 & C5.2\tabularnewline
\bottomrule[1.5pt]
\end{tabular}
    \caption{All flares produced by AR11967.\\
    $^{a}$Day of the observation of the strong field.}
    \label{tab:flares}
\end{table}

\section{Experiments with Multicomponent Atmospheres}\label{sec:experiments}
In Section \ref{sec:2comp} we described the experiments made with three
different models of multicomponent atmospheres. Table \ref{tab:atmos2comp}
summarizes the atmospheric conditions retrieved from the 2D
and 1D inversions for the reduced $\chi^2$, $|\Vec{B}|$,  $\gamma$, $v_{\rm LOS}$, and the filling factor $\alpha$ for five different pixels within the light bridge.

\end{appendix}

\bibliographystyle{aa}
\bibliography{references}

\begin{longrotatetable}
\begin{deluxetable*}{c>{\centering}m{.5cm}m{.5cm}m{.5cm}c*{19}{r}}
\setlength{\tabcolsep}{3pt}
\renewcommand{\arraystretch}{.8}
\tabletypesize{\footnotesize}
\tablehead{text}
\tablecaption{Atmospheric conditions retrieved from the 2D and 1D inversions for the comparison of one- and two-component inversions for selected pixels in Figures \ref{fig:2comp2d} and \ref{fig:2comp1d}.\label{tab:atmos2comp}}

\tablehead{
\multicolumn{4}{c}{} & \multicolumn{9}{c}{First Component} && \multicolumn{9}{c}{Second Component} \\
\cmidrule{5-13}
\cmidrule{15-24}
& \multirow{2}{*}{Pixel$^{a}$} & \multirow{2}{*}{dof$^{b}$}  &\multirow{2}{*}{$\chi^2$ $^{c}$} & \multicolumn{3}{c}{$|\Vec{B}|$ (kG)} & \multicolumn{3}{c}{$\gamma$ (deg)} & \multicolumn{3}{c}{$v_{\text{LOS}}$ (km/s)} &\multirow{2}{*}{$\alpha^{e}$} & \multicolumn{3}{c}{$|\Vec{B}|$ (kG)} & \multicolumn{3}{c}{$\gamma$ (deg)} & \multicolumn{3}{c}{$v_{\text{LOS}}$ (km/s)} \\
\cmidrule{5-13}\cmidrule{15-24}
\multicolumn{4}{c}{} & -2.3$^{d}$ &-0.8$^{d}$ & 0.0$^{d}$ & -2.3$^{d}$ &-0.8$^{d}$ & 0.0$^{d}$& -2.3$^{d}$ &-0.8$^{d}$ & 0.0$^{d}$ && -2.3$^{d}$ &-0.8$^{d}$ & 0.0$^{d}$ & -2.3$^{d}$ &-0.8$^{d}$ & 0.0$^{d}$& -2.3$^{d}$ &-0.8$^{d}$ & 0.0$^{d}$ & }
\startdata
\multirow{17}{*}{2D}& a &16 & 6.0 & 5.1 & 6.0 & 7.0 &114 &117 &125 & 0.8 & 8.8
& 4.1   & & & & & & & & & &\\
& b &16 & 12.5 & 4.1 & 4.6 & 5.9 &134 &138 &124 & 0.0 & 2.8 &12.4   & & & & & & &
& & &\\
& c &16 & 3.6 & 6.6 & 6.6 & 8.2 &122 &129 &151 & 5.1 & 6.6 & 3.0   & & & & & & &
& & &\\
& d &16 & 1.3 & 6.2 & 6.7 & 7.9 &123 &137 &150 & 1.4 & 3.0 & 4.2   & & & & & & &
& & &\\
& e &16 & 0.8 & 4.9 & 5.9 & 7.2 & 66 & 63 & 59 & 1.0 & 1.2 & 4.8   & & & & & & &
& & &\\
\cmidrule{2-23}
& a &24 & 1.5 & 5.7 & 6.3 & 5.5 &114 &118 &129 & 1.7 & 7.8 & 6.1 & 0.16
&& 2.4&& & 96&& &-1.3&    
        \\
& b &24 & 0.2 & 4.2 & 5.1 & 7.0 &138 &144 & 94 &-0.4 & 2.6 &13.5 & 0.09
&& 1.2&& &107&& & 4.7&   
        \\
& c &24 & 2.3 & 6.1 & 7.7 & 8.8 &122 &136 &157 & 6.3 & 5.1 & 4.9 & 0.18
&& 2.3&& & 83&& &-1.0&    
        \\
& d &24 & 2.3 & 5.9 & 6.7 & 8.0 &110 &136 &131 & 1.4 & 3.0 & 5.6 & 0.06
&& 2.4&& &106&& & 2.4&    
        \\
& e &24 & 1.7 & 5.2 & 6.0 & 7.2 & 71 & 60 & 55 & 0.9 & 1.5 & 2.8 & 0.12
&& 1.3&& & 86&& &-0.8&    
        \\
\cmidrule{2-23}
& a &32 & 1.3 & 5.2 & 6.3 & 6.5 &119 &117 &129 & 1.2 & 8.6 & 4.4 & 0.14 & 2.3
& 1.4 & 2.4 &  53.2 &  47.4 &  61.7 & -1.4 & -2.4 & -2.0 &\\
& b &32 & 0.3 & 3.3 & 5.6 & 6.0 &129 &149 &142 &-0.8 & 2.9 &10.1 & 0.08 & 1.3
& 1.4 & 4.8 &  47.8 &  70.6 & 123.1 &  3.3 & -0.5 & -4.0 &\\
& c &32 & 2.3 & 6.4 & 7.6 & 8.6 &128 &136 &152 & 4.0 & 5.3 & 5.1 & 0.21 & 1.9
& 3.4 & 3.9 &  70.0 & 137.5 & 138.4 &  7.2 &  1.5 &  2.2 &\\
& d &32 & 2.2 & 5.8 & 6.9 & 8.3 &116 &130 &143 & 0.7 & 2.6 & 6.2 & 0.04 & 1.4
& 2.0 & 2.2 & 154.6 & 120.3 &  70.2 &  5.5 &  0.2 & -0.8 &\\
& e &32 & 1.7 & 4.9 & 6.1 & 7.3 & 71 & 67 & 64 & 1.0 & 1.2 & 4.6 & 0.14 & 3.2
& 1.6 & 1.4 &  98.1 &  37.6 &  14.6 & -3.7 & -3.5 & -2.8 &\\
\midrule
\multirow{17}{*}{1D}& a &16 &10.9 & 5.0 & 6.0 & 6.5 &115 &120 &127 & 1.9 & 5.5
& 6.6   & & & & & & & & & &\\
& b &16 &42.7 & 3.6 & 4.4 & 4.8 &135 &139 &137 &-0.6 & 1.4 &11.7   & & & & & & &
& & &\\
& c &16 &15.5 & 5.4 & 6.6 & 7.1 &113 &126 &135 & 2.3 & 4.9 & 6.0   & & & & & & &
& & &\\
& d &16 & 4.7 & 5.7 & 6.7 & 7.3 &119 &124 &129 & 0.3 & 2.4 & 8.5   & & & & & & &
& & &\\
& e &16 & 2.2 & 4.7 & 6.1 & 7.0 & 75 & 72 & 64 & 0.6 & 1.4 & 6.3   & & & & & & &
& & &\\
\cmidrule{2-23}
& a &24 &9.1 & 5.3 & 5.9 & 6.2 &116 &121 &124 & 2.0 & 5.7 & 7.3 & 0.09
&& 1.8&& & 89&& &-0.7&    
        \\
& b &24 & 35.0 & 2.6 & 4.7 & 5.0 &127 &136 &121 &-1.6 & 1.3 &11.8 & 0.15
&& 1.4&& &144&& & 4.1&    
        \\
& c &24 & 9.7 & 5.7 & 7.2 & 7.8 &109 &126 &138 & 2.1 & 5.0 & 5.5 & 0.15
&& 3.5&& &161&& & 4.1&   
        \\
& d &24 & 3.7 & 5.2 & 6.4 & 7.3 &109 &122 &128 & 0.1 & 1.4 & 8.8 & 0.18
&& 3.1&& &153&& & 2.8&    
        \\
& e &24 & 1.4 & 3.7 & 5.7 & 7.2 & 78 & 74 & 63 & 1.0 & 0.8 & 5.5 & 0.37
&& 1.9&& & 70&& &-1.7&   
        \\
\cmidrule{2-23}
& a &32 &8.8 & 5.4 & 6.1 & 6.6 &116 &120 &137 & 1.9 & 5.2 & 6.7 & 0.09 & 1.9
& 2.6 & 2.7 &  61.6 & 129.0 & 151.1 & 11.5 &  4.0 &  1.4 &\\
& b &32 & 42.6 & 3.1 & 4.6 & 5.0 &126 &147 &173 &-0.7 & 1.1 &10.9 & 0.16 & 0.6
& 0.0 & 2.4 &   6.1 & 179.6 & 170.6 &  8.6 & -3.5 & -9.2 &\\
& c &32 & 7.9 & 5.5 & 7.3 & 7.6 &106 &121 &137 & 1.7 & 5.1 & 5.3 & 0.26 & 0.9
& 3.5 & 5.0 & 150.6 & 165.8 & 148.2 & 10.6 &  4.4 &  1.5 &\\
& d &32 & 3.6 & 5.7 & 6.5 & 7.2 &117 &126 &116 & 0.3 & 1.8 & 8.3 & 0.12 & 6.9
& 0.0 & 0.0 & 123.8 &  41.8 &   3.2 &  9.5 &  4.3 & -6.5 &\\
& e &32 & 1.1 & 4.0 & 5.6 & 6.9 & 78 & 74 & 63 & 0.7 & 0.6 & 5.7 & 0.35 & 2.8
& 2.3 & 2.5 &  81.5 &  71.7 &  84.3 &  2.1 & -4.5 & -8.7 &\\
\enddata
\tablecomments{t$^{a}${Pixels selected in Figures \ref{fig:2comp2d} and \ref{fig:2comp1d}.} $^{b}${Degrees of freedom of the inversion.} $^{c}${Reduced-$\chi^2$.} $^{d}${$\log\tau$.} $^{e}${Filling factor.}}
\end{deluxetable*}
\end{longrotatetable}

\end{document}